\documentclass[twocolumn,english,pra]{revtex4-1}
\usepackage[T1]{fontenc}
\usepackage[latin9]{inputenc}
\usepackage{color}
\usepackage{array}
\usepackage{multirow}
\usepackage{amsmath}
\usepackage{amssymb}
\usepackage{graphicx}

\makeatletter

\providecommand{\tabularnewline}{\\}

\usepackage{babel}

\makeatother

\usepackage{babel}
\begin{document}

\title{Criteria to detect genuine multipartite entanglement using spin measurements}

\author{R. Y. Teh and M. D. Reid}

\affiliation{$^{1}$Centre for Quantum and Optical Science Swinburne University
of Technology, Melbourne, Australia}
\begin{abstract}
We derive conditions in the form of inequalities to detect the full
$N$-partite inseparability and genuine $N$-partite entanglement
of $N$ systems. The inequalities are expressed in terms of variances
of spin operators, and can be tested by local spin measurements performed
on the individual systems. Violation of the inequalities is sufficient
(but not necessary) to certify the multipartite entanglement, and
occurs when a type of spin squeezing is created. The inequalities
are similar to those derived for continuous-variable systems, but
instead are based on the Heisenberg spin-uncertainty relation $\Delta J_{x}\Delta J_{y}\geq|\langle J_{z}\rangle|/2$.
We first derive criteria for full multipartite inseparability. Secondly,
we derive criteria for genuine multipartite entanglement in the limiting
case of large mean spin and reduced fluctuations, such that for each
system $\left|\langle J_{z}\rangle\right|$ is large and $\Delta^{2}J_{z}/\left|\langle J_{z}\rangle\right|\le1$.
Following that, we derive general criteria for genuine multipartite
entanglement without this assumption using a set of inequalities similar
to those considered for continuous-variable systems by van Loock and
Furusawa. We also extend previous work to derive spin-variance inequalities
that certify the full tripartite inseparability or genuine multi-partite
entanglement among systems with fixed spin $J$, as in Greenberger-Horne-Zeilinger
(GHZ) states and W states where $J=1/2$. These inequalities are derived
from the planar spin-uncertainty relation $(\Delta J_{x})^{2}+(\Delta J_{y})^{2}\geq C_{J}$
where $C_{J}$ is a constant for each $J$. Finally, it is shown how
the inequalities detect multipartite entanglement based on Stokes
operators. We illustrate with experiments that create entanglement
shared among separated atomic ensembles, polarization-entangled optical
modes, and the clouds of atoms of an expanding spin-squeezed Bose-Einstein
condensate.
\end{abstract}
\maketitle

\section{Introduction}

Genuine multipartite quantum entanglement is a resource required for
many protocols in the field of quantum information and computation
\cite{Gottesman_PRA1996,Bose_application_PRA1998,Hillery_application_PRA1999,vanLoock_application_PRL2000,Raussendorf_PRL2001,Jing_application_PRL2003,Giovannetti_application_Science2004,Briegel_application_Nature2009,Bancal_application_PRL2011}.\textcolor{red}{{}
}$N$ systems are said to be genuinely $N$-partite entangled if the
systems are mutually entangled in such a way that the entanglement
cannot be constructed by mixing entangled states involving fewer than
$N$ parties\textcolor{black}{{} \cite{Acin_biseparable_PRL2001,Jungnitsch_biseparable_PRL2011,Bancal_application_PRL2011}.
Mathematically, a tripartite system is genuinely tripartite entangled
if and only if the density operator characterizing the system cannot
be represented in the biseparable form \cite{horodecki_Phys_lett_A1996,Acin_biseparable_PRL2001,Jungnitsch_biseparable_PRL2011,Bancal_application_PRL2011}}\textcolor{red}{{}
} 
\begin{align}
\rho_{BS} & =P_{1}\sum_{R}\eta_{R}^{\left(1\right)}\rho_{1}^{R}\rho_{23}^{R}+P_{2}\sum_{R'}\eta_{R'}^{\left(2\right)}\rho_{2}^{R'}\rho_{13}^{R'}\nonumber \\
 & +P_{3}\sum_{R'}\eta_{R''}^{(3)}\rho_{3}^{R''}\rho_{12}^{R''}\,,\label{eq:biseparable_den_op}
\end{align}
where $\sum_{k=1}^{3}P_{k}=1$, $P_{k}\geq0$, and $\sum_{R}\eta_{R}^{\left(k\right)}=1$.
$\rho_{k}^{R}$ is an arbitrary density operator for the subsystem
$k$, while $\rho_{mn}^{R}$ is an arbitrary density operator for
the subsystems $m$ and $n$. The definition of genuine $N$-partite
entanglement follows similarly.

Criteria to certify genuine $N$-partite entanglement for continuous
variable (CV) systems have been derived by Shalm et al. \cite{Shalm_Nature2012}
and Teh and Reid \cite{PhysRevA.90.062337}. These criteria take the
form of variance inequalities, similar to those derived for CV bipartite
entanglement \cite{Reid_PRA1989,Duan_PRL2000,Giovannetti_PRA2003}.
The work of Refs. \cite{Shalm_Nature2012,PhysRevA.90.062337} extended
earlier work by van Loock and Furusawa, who developed CV criteria
for the related but different concept of full $N$-partite inseparability
\cite{vanLoock_PRA2003,Aoki_PRL2003} \textcolor{black}{(see also
Refs. \cite{Sperling_PRL2013,Gerke_PRX2018})}.{} Full $N$-partite
inseparability rules out the possibility that the system can be described
by a mixture of states of a given bipartition (so that only one of
the above $P_{i}$ can be nonzero), but does not rules out mixtures
of different bipartitions. Although genuine $N$-partite entanglement
implies full $N$-partite inseparability, the converse is not true,
and full $N$-partite inseparability is therefore a weaker form of
correlation. Nonetheless, for pure states, full $N$-partite inseparability
is sufficient to imply genuine $N$-partite entanglement. Experiments
have confirmed both full $N$-partite inseparability \cite{Aoki_PRL2003,Coelho_Science2009,Pfister_PRL2011,Armstrong_Nature2012,Gerke_Fabre_PRL2015}
 and genuine $N$-partite entanglement ($N\geq3$) for CV systems
\cite{Shalm_Nature2012,Armstrong_Nature2015,Barros_Walborn_CV_PRA2015,KunchiPeng_CV_Applied_Physics_Lett2016,SandboChang_CV_PRA2018}.
Here, ``continuous variable (CV)'' refers to the use of measurements
that have continuous-variable outcomes e.g. field quadrature phase
amplitudes $X$ and $P$, or position and momentum. The CV criteria
are derived from the commutation relation $\left[X,P\right]=2i$ and
the associated uncertainty relations.

In this paper, we derive criteria for genuine $N$-partite entanglement
and full $N$-partite inseparability that are useful for discrete
variable systems involving spin degrees of freedom. In this case,
measurements correspond to spin observables, and it is the spin commutation
relation $\left[J_{x},J_{y}\right]=iJ_{z}$ and associated spin-uncertainty
relations that are relevant. The criteria we derive involve variances
and apply to all physical systems, provided the measurements correspond
to operators satisfying spin commutation relation\textcolor{black}{s.
This approach extends to $N$ systems that of Hofmann and Takeuchi
\cite{Hofmann_PRA2003} and Raymer et al. \cite{PhysRevA.67.052104}
who used spin-uncertainty relations to derive variance criteria for
bipartite entanglement. The question of how to detect genuine $N$-partite
entanglement has been studied previously but most work has been in
the context of qubit (spin $1/2$) systems \cite{Svetlichny_PRD1987,Collins_PRL2002,Bourennane_PRL2004,Toth_PRA2005,Lougovski_NJP2009,Papp_Science2009,Lavoie_NJP2009,Horodecki_RMP2009,Guhne_NJP2010,Lu_tripartite_GHZPRA2011,He_Reid_steering_PRL2013}
or systems of fixed dimension \cite{Huber_highdimension_PRL2010,Huber_highdimension_PRA2011,Huber_highdimension_QIP2013,Li_highdimension_scientificreports2017,Regula_highdimension_NJP2018,Weinfurter_highdimension_EPJD2019}.}

\textcolor{black}{The development of criteria to certify the genuine
multipartite entanglement of discrete systems, as in this paper, is
motivated by the increasing number of experiments detecting entanglement
with atoms. For example, bipartite entanglement has been created between
atomic ensembles and separated atomic modes \cite{Julsgaard:2001aa,Fadel409,Lange416},
and multi-partite entanglement has been created among the separated
clouds of a Bose-Einstein condensate (BEC) \cite{Kunkel413}. It is
sometimes possible to rewrite the spin commutation relation in a form
that resembles the position-momentum commutation relation. This is
often true where the spin observables are expressed as Schwinger operators,
and justifies the use of CV entanglement criteria for the spin system
in that case. For instance, Julsgaard et al. \cite{Julsgaard:2001aa}
characterize the entanglement in the collective spins between two
atomic ensembles using CV criteria. However, as pointed out by Raymer
et al. \cite{PhysRevA.67.052104}, this is only valid in a restricted
sense and will not give correct results in general. In other words,
the complete spin commutation relation should be used in any derivation
of criteria certifying the genuine multipartite entanglement of spin
systems.}

\textcolor{black}{The program of characterizing entanglement in spin
systems has been largely motivated by the observation that a spin-squeezed
system exhibits quantum correlations among the spin particles. S{ø}rensen
et al. \cite{Sorensen_Nature2001} derived an $N$-partite entanglement
criterion that implies the presence of an $N$-partite entangled state.
Here, an $N$-partite entangled state is a state that cannot be expressed
in the form 
\begin{align}
\rho_{S} & =\sum_{R}P_{R}\rho_{R}^{(1)}\rho_{R}^{(2)}...\rho_{R}^{(N)}\,,\label{eq:N_separable}
\end{align}
where $\sum_{R}P_{R}=1$. A host of criteria \cite{Toth_PRA2004,Toth_PRL2007,Toth_JOSAB2007,Toth_PRA2009,Vitagliano_PRL2011,He_Drummond_Reid_higher_spin_PRA2011}
were subsequently derived to certify the presence of $N$-partite
entanglement in spin systems. However, these criteria only rule out
the possibility of $N$-partite separable states of the form Eq. (\ref{eq:N_separable})
and not the more general $N$-partite biseparable states of the form
Eq. (\ref{eq:biseparable_den_op}) (as extended to higher $N$) where
all separable bipartitions (and mixtures of them) are considered.
Hence they are not criteria for genuine $N$-partite entanglement,
where the entanglement is mutually shared among all $N$ parties.
An exception is the spin-squeezing criteria of S{ø}rensen and M{ø}lmer
(and others like it) which imply a genuine $k$-particle entanglement
shared among $k$ particles of an $N$-particle system ($k\leq N$)
\cite{Sorensen_spin_squeezing_RPL2001}. Such criteria differ from
those derived in this paper, however, being based on collective spin
measurements made on the composite system, rather than local measurements
made on separated subsystems, and thus cannot directly test nonlocal
models (as described in Ref. \cite{Cavalcanti_PRA2012}).}

\textcolor{black}{The task of characterizing genuine multipartite
entanglement in spin systems was carried out by Korbicz et al. \cite{Korbicz_PRL2005,Korbicz_PRA2006}.
Korbicz and co-workers used the positivity of partial transpose (PPT)
criterion or the Peres-Horodecki criterion \cite{Peres_PRL1996,horodecki_Phys_lett_A1996}
as the starting point to derive entanglement criteria, and showed
genuine tripartite entanglement for symmetric states. The PPT criterion,
however, is less useful for $N$-partite separability when $N$ is
large \cite{horodecki_Phys_lett_A1996}. In this paper, we derive
criteria for genuine multipartite entanglement for spin systems by
ruling out the possibility of the state in a biseparable form as in
Eq. (\ref{eq:biseparable_den_op}) using the uncertainty relations
for spin operators.}

The remainder of the paper is structured as follows. In Section \ref{sec:tripartite},
we derive criteria to detect the full tripartite inseparability and
genuine tripartite entanglement of three systems using spin measurements.
The generalization to genuine $N$-partite entanglement is given in
Section \ref{sec:N_partite}. These criteria are derived using methods
similar to those developed by van Loock and Fursusawa \cite{vanLoock_PRA2003},
Teh and Reid \cite{PhysRevA.90.062337} and Shalm et al. \cite{Shalm_Nature2012}
for CV systems. In Section III, we extend variance criteria derived
for Einstein-Podolsky-Rosen (EPR) steering by He and Reid \cite{He_Reid_steering_PRL2013},
pointing out that these inequalities apply to certify genuine tripartite
entanglement as well as steering, which is a form of entanglement
closely connected with the EPR paradox\textcolor{black}{{} \cite{Wiseman_PRL2007,Jones_PRA2007,Cavalcanti_PRA2012}.}
The criteria are derived using planar spin-uncertainty relations \cite{PhysRevA.84.022107,He_planar_squeeze_NJP2012,Vitagliano_spin_squeezing_NJP2017,Vitagliano_spin_squeezing_PRA2018,Puentes_planar_squeeze_NJP2013}
and apply to subsystems with a fixed spin $J$. We show that the criteria
may be used to detect the genuine tripartite entanglement of GHZ states,
and the full tripartite inseparability of W states. Finally, in Section
\ref{sec:Applications}, we explain how to generate genuinely-entangled
spin systems based on Stokes operators. We then demonstrate the application
of the criteria derived in Sections II and IV to certify full $N$-partite
inseparability and genuine $N$-partite entanglement of these systems.

\section{Criteria for full tripartite inseparability and genuine tripartite
entanglement\label{sec:tripartite}}

The criteria derived in this section involve variances of the sum
of spin observables defined for each subsystem. In this work, all
the caret symbols that denote the spin operators are dropped, unless
specified otherwise, and we use the symbol $\Delta^{2}x$ to denote
the variance of $x$.

In this section, we will derive criteria for both full tripartite
inseparability and genuine tripartite entanglement. Criteria 1 and
3 will be applicable to full tripartite inseparability, but only
apply to genuine tripartite entanglement in a limit of large spin.
Criteria 2 and 4 are applicable to both full tripartite inseparability
and genuine tripartite entanglement, for states with arbitrary spin.

\subsection{The sum inequalities \label{subsec:The-sum-inequalities}}

\subsubsection{Sum of two variances}

Consider the sum of $\Delta^{2}u$ and $\Delta^{2}v$ where 
\begin{eqnarray}
u & = & h_{1}J_{x,1}+h_{2}J_{x,2}+h_{3}J_{x,3}\nonumber \\
v & = & g_{1}J_{y,1}+g_{2}J_{y,2}+g_{3}J_{y,3}\label{eq:uv}
\end{eqnarray}
and $h_{k}$ and $g_{k}$ ($k=1,2,3$) are real numbers. Here, $J_{x,k},J_{y,k}$,
$J_{z,k}$ are the spin operators for subsystem $k$, satisfying the
commutation relation $\left[J_{x,k},J_{y,k}\right]=iJ_{z,k}$. We
derive the bound for $\Delta^{2}u+\Delta^{2}v$ such that the violation
of the bound implies the genuine tripartite entanglement in the spin
degree of freedom.\textcolor{red}{{} }

\textcolor{black}{This leads us to Criterion 1. }Firstly, we assume
that the spin state is in a biseparable mixture state 
\begin{align*}
\rho_{BS} & =P_{1}\sum\eta_{R}^{(1)}\rho_{1}^{R}\rho_{23}^{R}+P_{2}\sum_{R'}\eta_{R'}^{(2)}\rho_{13}^{R'}\rho_{2}^{R'}\\
 & +P_{3}\sum_{R''}\eta_{R''}^{(3)}\rho_{12}^{R''}\rho_{3}^{R''}
\end{align*}
as in Eq. (\ref{eq:biseparable_den_op}). This implies that the variance
of an observable $\Delta^{2}u$ is greater or equal to the sum of
the variances of the observable of its component state $\Delta^{2}u_{R}$,
i.e. \cite{Hofmann_PRA2003} 
\begin{align*}
\Delta^{2}u & \geq\sum_{R}P_{R}\Delta^{2}u_{R}\,.
\end{align*}
The sum of $\Delta^{2}u$ and $\Delta^{2}v$ is then 
\begin{align}
\Delta^{2}u+\Delta^{2}v & \geq P_{1}\sum_{R}\eta_{R}^{(1)}\left[\Delta^{2}u_{R}+\Delta^{2}v_{R}\right]\nonumber \\
 & +P_{2}\sum_{R'}\eta_{R'}^{(2)}\left[\Delta^{2}u_{R'}+\Delta^{2}v_{R'}\right]\nonumber \\
 & +P_{3}\sum_{R''}\eta_{R''}^{(3)}\left[\Delta^{2}u_{R''}+\Delta^{2}v_{R''}\right]\,.\label{eq:var_uplusvar_v}
\end{align}
To proceed, we consider $\Delta^{2}u_{\zeta}+\Delta^{2}v_{\zeta}$
that corresponds to an arbitrary bipartition $\rho_{k}^{\zeta}\rho_{lm}^{\zeta}$:
\begin{eqnarray}
\Delta^{2}u_{\zeta}+\Delta^{2}v_{\zeta} & \geq & \left|g_{k}h_{k}\langle J_{z,k}\rangle_{\zeta}\right|\!\nonumber \\
 &  & +\!\left|g_{l}h_{l}\langle J_{z,l}\rangle_{\zeta}+g_{m}h_{m}\langle J_{z,m}\rangle_{\zeta}\right|\,.\label{eq:varu_plus_varv_arbitrary}
\end{eqnarray}
\textcolor{black}{The lower bound given in this inequality is derived
in the Appendix 1, using the uncertainty relations for spin. }With
these lower bounds for different bipartitions, Eq. (\ref{eq:var_uplusvar_v})
has the expression
\begin{widetext}
\vspace{-0.5mm}
\begin{align}
\Delta^{2}u+\Delta^{2}v & \geq\!P_{1}\sum_{R}\eta_{R}^{(1)}\!\left|g_{1}h_{1}\langle J_{z,1}\rangle_{R}+g_{2}h_{2}\langle J_{z,2}\rangle_{R}+g_{3}h_{3}\langle J_{z,3}\rangle_{R}\right|\!+\!P_{2}\!\sum_{R'}\eta_{R'}^{(2)}\left|g_{2}h_{2}\langle J_{z,2}\rangle_{R'}+g_{1}h_{1}\langle J_{z,1}\rangle_{R'}+g_{3}h_{3}\langle J_{z,3}\rangle_{R'}\right|\nonumber \\
 & +P_{3}\sum_{R''}\eta_{R''}^{(3)}\left|g_{3}h_{3}\langle J_{z,3}\rangle_{R''}+g_{1}h_{1}\langle J_{z,1}\rangle_{R''}+g_{2}h_{2}\langle J_{z,2}\rangle_{R''}\right|\,\label{eq:gen}
\end{align}
\vspace{-0.5mm}
\end{widetext}

\textbf{\emph{Criterion 1\label{1}.}}\textbf{ }Violation of the inequality
\begin{align}
\Delta^{2}u+\Delta^{2}v & \geq\text{min}\left\{ \left|g_{1}h_{1}\langle J_{z,1}\rangle\right|+\left|g_{2}h_{2}\langle J_{z,2}\rangle+g_{3}h_{3}\langle J_{z,3}\rangle\right|\,,\right.\nonumber \\
 & \left|g_{2}h_{2}\langle J_{z,2}\rangle\right|+\left|g_{1}h_{1}\langle J_{z,1}\rangle+g_{3}h_{3}\langle J_{z,3}\rangle\right|\,,\nonumber \\
 & \left.\left|g_{3}h_{3}\langle J_{z,3}\rangle\right|+\left|g_{1}h_{1}\langle J_{z,1}\rangle+g_{2}h_{2}\langle J_{z,2}\rangle\right|\right\} \,\label{eq:criterion-1}
\end{align}
is sufficient to confirm full tripartite inseparability. Genuine tripartite
entanglement is confirmed if the inequality is violated in the limit
where the mean spins are large, $\left|\langle J_{z,k}\rangle\right|\rightarrow\infty$,
and where fluctuations satisfy $\Delta^{2}J_{z,k}/\left|\langle J_{z,k}\rangle\right|\le1$.

\emph{Proof}. To prove the first statement, we need to rule out all
any given fixed bipartition i.e. we need to exclude (\ref{eq:biseparable_den_op})
where only one of the $P_{i}$ can be nonzero. This follows straightforwardly,
from (\ref{eq:var_uplusvar_v}) on using (\ref{eq:varu_plus_varv_arbitrary})
where only one of the $P_{1}$, $P_{2,}$ or $P_{3}$ can be nonzero.
Taking the minimum value (\ref{eq:varu_plus_varv_arbitrary}) over
the three possible bipartitions gives the required result.

To prove the second result, we first note that 
\begin{eqnarray}
\Delta^{2}J_{z,k} & \geq & P_{1}\langle(J_{z,k}-\langle J_{z,k}\rangle_{I})^{2}\rangle_{I}\nonumber \\
 &  & +P_{2}\langle(J_{z,k}-\langle J_{z,k}\rangle_{II})^{2}\rangle_{II}\nonumber \\
 &  & +P_{3}\langle(J_{z,k}-\langle J_{z,k}\rangle_{III})^{2}\rangle_{III}\,,\label{eq:11}
\end{eqnarray}
where we define $\langle J_{z,k}\rangle_{I}=\sum_{R}\eta_{R}^{(1)}\langle J_{z,k}\rangle_{R}$,
$\langle J_{z,k}\rangle_{II}=\sum_{R'}\eta_{R'}^{(1)}\langle J_{z,k}\rangle_{R'}$
and $\langle J_{z,k}\rangle_{III}=\sum_{R''}\eta_{R''}^{(1)}\langle J_{z,k}\rangle_{R''}$.
The subscripts $I$, $II$ and $III$ indicate averages and variances
over all the states with the given bipartitions $\rho_{1}\rho_{23}$,
$\rho_{2}\rho_{13}$ and $\rho_{3}\rho_{12}$, respectively. We thus
obtain\begin{widetext}

\begin{flalign*}
\frac{1}{|\langle J_{z,k}\rangle|^{2}}\{P_{1}\langle(J_{z,k}-\langle J_{z,k}\rangle_{I})^{2}\rangle_{I}+P_{2}\langle(J_{z,k}-\langle J_{z,k}\rangle_{II})^{2}\rangle_{II}+P_{3}\langle(J_{z,k}-\langle J_{z,k}\rangle_{III})^{2}\rangle_{III}\} & \leq\frac{\Delta^{2}J_{z,k}}{|\langle J_{z,k}\rangle|^{2}}\rightarrow0
\end{flalign*}
\end{widetext}as $|\langle J_{z,k}\rangle|\rightarrow\infty$. This
also implies, from Eq. (\ref{eq:11}), that $\frac{\langle J_{z,k}\rangle_{I}}{\langle J_{z,k}\rangle}=\frac{\langle J_{z,k}\rangle_{II}}{\langle J_{z,k}\rangle}=\frac{\langle J_{z,k}\rangle_{III}}{\langle J_{z,k}\rangle}\rightarrow1.$
Continuing from Eq. (\ref{eq:gen}), $\Delta^{2}u+\Delta^{2}v$ satisfies
the following inequality: 
\begin{widetext}
\begin{eqnarray}
\Delta^{2}u+\Delta^{2}v & \geq & P_{1}\left(\left|g_{1}h_{1}\langle J_{z,1}\rangle\frac{\langle J_{z,1}\rangle{}_{I}}{\langle J_{z,1}\rangle}\right|+\left|g_{2}h_{2}\langle J_{z,2}\rangle\frac{\langle J_{z,2}\rangle_{I}}{\langle J_{z,2}\rangle}+g_{3}h_{3}\langle J_{z,3}\rangle\frac{\langle J_{z,3}\rangle{}_{I}}{\langle J_{z,3}\rangle}\right|\right)\nonumber \\
 &  & +P_{2}\left(\left|g_{2}h_{2}\langle J_{z,2}\rangle\frac{\langle J_{z,2}\rangle{}_{II}}{\langle J_{z,2}\rangle}\right|+\left|g_{1}h_{1}\langle J_{z,1}\rangle\frac{\langle J_{z,1}\rangle_{II}}{\langle J_{z,1}\rangle}+g_{3}h_{3}\langle J_{z,3}\rangle\frac{\langle J_{z,3}\rangle{}_{II}}{\langle J_{z,3}\rangle}\right|\right)\nonumber \\
 &  & +P_{3}\left(\left|g_{3}h_{3}\langle J_{z,3}\rangle\frac{\langle J_{z,3}\rangle{}_{III}}{\langle J_{z,3}\rangle}\right|+\left|g_{2}h_{2}\langle J_{z,2}\rangle\frac{\langle J_{z,2}\rangle_{III}}{\langle J_{z,2}\rangle}+g_{1}h_{1}\langle J_{z,1}\rangle\frac{\langle J_{z,1}\rangle{}_{III}}{\langle J_{z,1}\rangle}\right|\right)\nonumber \\
 & \rightarrow & P_{1}\left(\left|g_{1}h_{1}\langle J_{z,1}\rangle\right|+\left|g_{2}h_{2}\langle J_{z,2}\rangle+g_{3}h_{3}\langle J_{z,3}\rangle\right|\right)+P_{2}\left(\left|g_{2}h_{2}\langle J_{z,2}\rangle\right|+\left|g_{1}h_{1}\langle J_{z,1}\rangle+g_{3}h_{3}\langle J_{z,3}\rangle\right|\right)\nonumber \\
 &  & +P_{3}\left(\left|g_{3}h_{3}\langle J_{z,3}\rangle\right|+\left|g_{2}h_{2}\langle J_{z,2}\rangle+g_{1}h_{1}\langle J_{z,1}\rangle\right|\right)\,.\label{eq:large_spin}
\end{eqnarray}
We can always choose for the lower bound the smallest value of the
terms in bracket in Eq. (\ref{eq:large_spin}). Hence, Eq. (\ref{eq:large_spin})
becomes Eq. (\ref{eq:criterion-1}), where we use the fact that $\sum\eta_{R}^{(1)}=1$
and $\sum P_{k}=1$. $\square$
\end{widetext}

In Eq. (\ref{eq:criterion-1}), the first term in the bracket $\{\}$
is implied by the biseparable state $\rho_{1}\rho_{23}$, the second
term is implied by the biseparable state $\rho_{2}\rho_{13}$, and
the final term is implied by the biseparable state $\rho_{3}\rho_{12}$.
We have taken large mean spins and small fluctuations, such that $\Delta^{2}J_{z,k}/\left|\langle J_{z,k}\rangle\right|\leq1$
and hence $\Delta J_{z,k}/|\langle J_{z,k}\rangle|\rightarrow0$ as
$|\langle J_{z,k}\rangle|\rightarrow\infty$. The optimal values for
$g_{k},h_{k}$ depend on the specific spin state. The criterion given
by Eq. (\ref{eq:criterion-1}) is thus a general result that allows
us to derive a host of other criteria. Examples of optimal choices
for different types of spin states will be given in Section \ref{sec:Applications}.
We derive further criteria for genuine tripartite entanglement without
the limiting assumptions in the next subsection.

\subsubsection{Van Loock-Furusawa inequalities for spin}

We can also derive the spin version of a set of inequalities derived
by van Loock and Furusawa \cite{vanLoock_PRA2003}. The quantities
$B_{I}$, $B_{II}$ and $B_{III}$ are defined as 
\begin{align}
B_{I} & \equiv\Delta^{2}\left(J_{x,1}-J_{x,2}\right)+\Delta^{2}\left(J_{y,1}+J_{y,2}+g_{3}J_{y,3}\right)\nonumber \\
B_{II} & \equiv\Delta^{2}\left(J_{x,2}-J_{x,3}\right)+\Delta^{2}\left(g_{1}J_{y,1}+J_{y,2}+J_{y,3}\right)\nonumber \\
B_{III} & \equiv\Delta^{2}\left(J_{x,1}-J_{x,3}\right)+\Delta^{2}\left(J_{y,1}+g_{2}J_{y,2}+J_{y,3}\right)\,.\label{eq:vlf_B}
\end{align}
Criteria can be derived both in the limiting case $\Delta J_{z,k}/|\langle J_{z,k}\rangle|\rightarrow0$
as $|\langle J_{z,k}\rangle|\rightarrow\infty$, considered in Criterion
1, and more generally, as we will show in Criterion 2. In the limiting
case considered above, the inequalities can be used to demonstrate
tripartite inseparability and genuine tripartite entanglement. We
show this next.

Considering $\left|\langle J_{z,k}\rangle\right|\rightarrow\infty$
and $\Delta^{2}J_{z,k}/\left|\langle J_{z,k}\rangle\right|\le1$,
the violation of any two of the inequalities in the set of inequalities
below:

\begin{align}
B_{I} & \geq\left(\left|\langle J_{z,1}\rangle\right|+\left|\langle J_{z,2}\rangle\right|\right)\nonumber \\
B_{II} & \geq\left(\left|\langle J_{z,2}\rangle\right|+\left|\langle J_{z,3}\rangle\right|\right)\nonumber \\
B_{III} & \geq\left(\left|\langle J_{z,1}\rangle\right|+\left|\langle J_{z,3}\rangle\right|\right)\,\label{eq:vlf_inseparability}
\end{align}
implies full tripartite inseparability.\medskip{}

\emph{Proof}. Note that for a biseparable state $\rho_{1}\rho_{23}=\sum_{R}\eta_{R}^{(1)}\rho_{1}^{R}\rho_{23}^{R}$,
the inequality $\Delta^{2}u+\Delta^{2}v\geq\left|g_{1}h_{1}\langle J_{z,1}\rangle_{R}\right|+\left|g_{2}h_{2}\langle J_{z,2}\rangle_{R}+g_{3}h_{3}\langle J_{z,3}\rangle_{R}\right|$
holds for a tripartite separable state, while the inequality $\Delta^{2}u+\Delta^{2}v\geq\left|g_{2}h_{2}\langle J_{z,2}\rangle_{R'}\right|+\left|g_{1}h_{1}\langle J_{z,1}\rangle_{R'}+g_{3}h_{3}\langle J_{z,3}\rangle_{R'}\right|$
is satisfied by the biseparable state $\rho_{2}\rho_{13}=\sum_{R'}\eta_{R'}^{(2)}\rho_{2}^{R'}\rho_{13}^{R'}$,
and the biseparable state $\rho_{3}\rho_{12}=\sum_{R''}\eta_{R''}^{(3)}\rho_{3}^{R''}\rho_{12}^{R''}$
satisfies $\Delta^{2}u+\Delta^{2}v\geq\left|g_{3}h_{3}\langle J_{z,3}\rangle_{R''}\right|+\left|g_{1}h_{1}\langle J_{z,1}\rangle_{R''}+g_{2}h_{2}\langle J_{z,2}\rangle_{R''}\right|$.
In the limit of $\left|\langle J_{z,k}\rangle\right|\rightarrow\infty$
and $\Delta^{2}J_{z,k}/\left|\langle J_{z,k}\rangle\right|\le1$,
the derivation as in Eq. (\ref{eq:large_spin}) leads to measurable
bounds. The inequalities that are satisfied by $\rho_{1}\rho_{23}$,
$\rho_{2}\rho_{13}$ and $\rho_{3}\rho_{12}$ are respectively
\begin{align}
\Delta^{2}u+\Delta^{2}v & \geq\left|g_{1}h_{1}\langle J_{z,1}\rangle\right|+\left|g_{2}h_{2}\langle J_{z,2}\rangle+g_{3}h_{3}\langle J_{z,3}\rangle\right|\nonumber \\
\Delta^{2}u+\Delta^{2}v & \geq\left|g_{2}h_{2}\langle J_{z,2}\rangle\right|+\left|g_{1}h_{1}\langle J_{z,1}\rangle+g_{3}h_{3}\langle J_{z,3}\rangle\right|\nonumber \\
\Delta^{2}u+\Delta^{2}v & \geq\left|g_{3}h_{3}\langle J_{z,3}\rangle\right|+\left|g_{1}h_{1}\langle J_{z,1}\rangle+g_{2}h_{2}\langle J_{z,2}\rangle\right|.\label{eq:vlf_largespin}
\end{align}

By choosing the coefficients $g_{k}$ and $h_{k}$ in Eq. (\ref{eq:vlf_largespin}),
we obtain a set of inequalities satisfied by $B_{I}$, $B_{II}$ and
$B_{III}$. For example, the left side of the criterion in Eq. (\ref{eq:vlf_largespin})
is equal to $B_{I}$ when $h_{1}=1,\,h_{2}=-1,\,h_{3}=0$ and $g_{1}=g_{2}=1$.
This gives 
\begin{align*}
B_{I} & \geq\left|\langle J_{z,1}\rangle\right|+\left|\langle J_{z,2}\rangle\right|\,,
\end{align*}
which is implied by the biseparable states $\rho_{1}\rho_{23}$ and
$\rho_{2}\rho_{13}$. Similarly choosing $h_{2}=g_{2}=g_{3}=-h_{3}=1$
and $h_{1}=0$, we obtain $B_{II}\geq\left(\left|\langle J_{z,2}\rangle\right|+\left|\langle J_{z,3}\rangle\right|\right)$
which is implied by the biseparable states $\rho_{2}\rho_{13}$ and
$\rho_{3}\rho_{12}$. Finally, $B_{III}\geq\left(\left|\langle J_{z,1}\rangle\right|+\left|\langle J_{z,3}\rangle\right|\right)$
is satisfied by the biseparable states $\rho_{1}\rho_{23}$ and $\rho_{3}\rho_{12}$.
Thus, in the limiting case $\left|\langle J_{z,k}\rangle\right|\rightarrow\infty$
and $\Delta^{2}J_{z,k}/\left|\langle J_{z,k}\rangle\right|\le1$,
full tripartite inseparability can be proved if any two of the inequalities
(\ref{eq:vlF_sum}) are violated. $\square$

The full tripartite inseparability does not rule out the possibility
of a biseparable mixture state, and thus, does not signify genuine
tripartite entanglement. When the biseparable mixture state is taken
into account, the van Loock-Furusawa set of inequalities is given
below: 
\begin{align}
B_{I} & \geq\left(P_{1}+P_{2}\right)\left(\left|\langle J_{z,1}\rangle\right|+\left|\langle J_{z,2}\rangle\right|\right)\nonumber \\
B_{II} & \geq\left(P_{2}+P_{3}\right)\left(\left|\langle J_{z,2}\rangle\right|+\left|\langle J_{z,3}\rangle\right|\right)\nonumber \\
B_{III} & \geq\left(P_{1}+P_{3}\right)\left(\left|\langle J_{z,1}\rangle\right|+\left|\langle J_{z,3}\rangle\right|\right)\,.\label{eq:vlF_sum}
\end{align}
This leads to a condition for genuine tripartite inseparability. Genuine
tripartite entanglement is confirmed if the inequality 
\begin{equation}
B_{I}+B_{II}+B_{III}\geq\left|\langle J_{z,1}\rangle\right|+\left|\langle J_{z,2}\rangle\right|+\left|\langle J_{z,3}\rangle\right|\,\label{eq:crit}
\end{equation}
is violated.\vskip0.1in

\emph{Proof. }It is first useful to prove this in the limiting case
where with $\left|\langle J_{z,k}\rangle\right|\rightarrow\infty$
and $\Delta^{2}J_{z,k}/\left|\langle J_{z,k}\rangle\right|\le1$.
By choosing $h_{1}=g_{1}=-h_{2}=g_{2}=1$ and $h_{3}=0$, inequality
Eq. (\ref{eq:gen}) becomes 
\begin{align}
B_{I} & \geq P_{1}\sum_{R}\eta_{R}^{(1)}\left[\left|\langle J_{z,1}\rangle_{R}\right|+\left|\langle J_{z,2}\rangle_{R}\right|\right]\nonumber \\
 & +P_{2}\sum_{R'}\eta_{R'}^{(2)}\left[\left|\langle J_{z,1}\rangle_{R'}\right|+\left|\langle J_{z,2}\rangle\right|_{R'}\right]\nonumber \\
 & \rightarrow\left(P_{1}+P_{2}\right)\left(\left|\langle J_{z,1}\rangle\right|+\left|\langle J_{z,2}\rangle\right|\right)\,,\label{eq:B_I-1}
\end{align}
where the final line is obtained using the same limit as in Eq. (\ref{eq:large_spin})
with $\left|\langle J_{z,k}\rangle\right|\rightarrow\infty$ and $\Delta^{2}J_{z,k}/\left|\langle J_{z,k}\rangle\right|\le1$.
Similarly choosing $h_{2}=g_{2}=g_{3}=-h_{3}=1$ and $h_{1}=0$ 
\begin{align}
B_{II} & \equiv\Delta^{2}\left(J_{x,2}-J_{x,3}\right)+\Delta^{2}\left(g_{1}J_{y,1}+J_{y,2}+J_{y,3}\right)\nonumber \\
 & \geq P_{2}\sum_{R'}\eta_{R'}^{(2)}\left[\left|\langle J_{z,2}\rangle_{R'}\right|+\left|\langle J_{z,3}\rangle\right|_{R'}\right]\nonumber \\
 & +P_{3}\sum_{R''}\eta_{R''}^{(3)}\left[\left|\langle J_{z,2}\rangle_{R''}\right|+\left|\langle J_{z,3}\rangle\right|_{R''}\right]\nonumber \\
 & \rightarrow\left(P_{2}+P_{3}\right)\left(\left|\langle J_{z,2}\rangle\right|+\left|\langle J_{z,3}\rangle\right|\right)\label{eq:B_II_1}
\end{align}
in the same limit. Finally, 
\begin{align}
B_{III} & \equiv\Delta^{2}\left(J_{x,1}-J_{x,3}\right)+\Delta^{2}\left(J_{y,1}+g_{2}J_{y,2}+J_{y,3}\right)\nonumber \\
 & \geq P_{1}\sum_{R}\eta_{R}^{(1)}\left[\left|\langle J_{z,1}\rangle_{R}\right|+\left|\langle J_{z,3}\rangle\right|_{R}\right]\nonumber \\
 & +P_{3}\sum_{R''}\eta_{R''}^{(3)}\left[\left|\langle J_{z,1}\rangle_{R''}\right|+\left|\langle J_{z,3}\rangle\right|_{R''}\right]\,\nonumber \\
 & \rightarrow\left(P_{1}+P_{3}\right)\left(\left|\langle J_{z,1}\rangle\right|+\left|\langle J_{z,3}\rangle\right|\right)\,.\label{eq:B_III_1}
\end{align}
Using the inequalities in Eq. (\ref{eq:vlF_sum}), we can readily
prove the criterion for genuine tripartite entanglement in the limiting
case. Taking the sum of $B_{I}$, $B_{II}$ and $B_{III}$ in Eqs.
(\ref{eq:B_I-1}), (\ref{eq:B_II_1}) and (\ref{eq:B_III_1}) respectively,
we obtain 
\begin{align*}
B_{I}+B_{II}+B_{III} & \geq\left(2P_{1}+P_{2}+P_{3}\right)\left|\langle J_{z,1}\rangle\right|\\
 & +\left(P_{1}+2P_{2}+P_{3}\right)\left|\langle J_{z,2}\rangle\right|\\
 & +\left(P_{1}+P_{2}+2P_{3}\right)\left|\langle J_{z,3}\rangle\right|
\end{align*}
Continuing, we see that
\begin{align*}
B_{I}+B_{II}+B_{III} & \geq\left(P_{1}+P_{2}+P_{3}\right)\\
 & \ \ \ \times\left(\left|\langle J_{z,1}\rangle\right|+\left|\langle J_{z,2}\rangle\right|+\left|\langle J_{z,3}\rangle\right|\right)\\
 & =\left(\left|\langle J_{z,1}\rangle\right|+\left|\langle J_{z,2}\rangle\right|+\left|\langle J_{z,3}\rangle\right|\right)\,.
\end{align*}
$\square$

The condition for genuine tripartite entanglement can be proved for
all parameters, \emph{without} the assumptions $\left|\langle J_{z,k}\rangle\right|\rightarrow\infty$
and $\Delta^{2}J_{z,k}/\left|\langle J_{z,k}\rangle\right|\le1$.
We do this below.\bigskip{}

\textbf{\textit{\textcolor{black}{Criterion 2\label{2}. }}}\textit{\textcolor{black}{\emph{For
arbitrary states, both full tripartite inseparability and genuine
tripartite entanglement are observed if the inequality 
\begin{align}
B_{I}+B_{II}+B_{III} & \geq\left|\langle J_{z,1}\rangle\right|+\left|\langle J_{z,2}\rangle\right|+\left|\langle J_{z,3}\rangle\right|\,\label{eq:genuine_tripartite_vlf_sums}
\end{align}
is violated.}}}

\textit{Proof}. By choosing $h_{1}=g_{1}=-h_{2}=g_{2}=1$ and $h_{3}=0$,
inequality Eq. (\ref{eq:gen}) becomes 
\begin{align}
B_{I} & \geq P_{1}\sum_{R}\eta_{R}^{(1)}\left[\left|\langle J_{z,1}\rangle_{R}\right|+\left|\langle J_{z,2}\rangle_{R}\right|\right]\nonumber \\
 & +P_{2}\sum_{R'}\eta_{R'}^{(2)}\left[\left|\langle J_{z,1}\rangle_{R'}\right|+\left|\langle J_{z,2}\rangle\right|_{R'}\right]\,\label{eq:B-1}
\end{align}
since the term involving $P_{3}$ is always non-negative. Similarly
choosing $h_{2}=g_{2}=g_{3}=-h_{3}=1$ and $h_{1}=0$ 
\begin{align}
B_{II} & \equiv\Delta^{2}\left(J_{x,2}-J_{x,3}\right)+\Delta^{2}\left(g_{1}J_{y,1}+J_{y,2}+J_{y,3}\right)\nonumber \\
 & \geq P_{2}\sum_{R'}\eta_{R'}^{(2)}\left[\left|\langle J_{z,2}\rangle_{R'}\right|+\left|\langle J_{z,3}\rangle\right|_{R'}\right]\nonumber \\
 & +P_{3}\sum_{R''}\eta_{R''}^{(3)}\left[\left|\langle J_{z,2}\rangle_{R''}\right|+\left|\langle J_{z,3}\rangle\right|_{R''}\right]\label{eq:B-2}
\end{align}
Similarly, 
\begin{align}
B_{III} & \equiv\Delta^{2}\left(J_{x,1}-J_{x,3}\right)+\Delta^{2}\left(J_{y,1}+g_{2}J_{y,2}+J_{y,3}\right)\nonumber \\
 & \geq P_{1}\sum_{R}\eta_{R}^{(1)}\left[\left|\langle J_{z,1}\rangle_{R}\right|+\left|\langle J_{z,3}\rangle\right|_{R}\right]\nonumber \\
 & +P_{3}\sum_{R''}\eta_{R''}^{(3)}\left[\left|\langle J_{z,1}\rangle_{R''}\right|+\left|\langle J_{z,3}\rangle\right|_{R''}\right]\,\label{eq:B-3}
\end{align}
From these inequalities, we now prove Criterion 2 for genuine tripartite
entanglement:
\begin{widetext}
\begin{eqnarray}
B_{I}+B_{II}+B_{III} & \geq & P_{1}\sum_{R}\eta_{R}^{(1)}\left(2\left|\langle J_{z,1}\rangle_{R}\right|+\left|\langle J_{z,2}\rangle_{R}\right|+\left|\langle J_{z,3}\rangle_{R}\right|\right)+P_{2}\sum_{R'}\eta_{R'}^{(2)}\left(\left|\langle J_{z,1}\rangle_{R'}\right|+2\left|\langle J_{z,2}\rangle_{R'}\right|+\left|\langle J_{z,3}\rangle_{R'}\right|\right)\nonumber \\
 &  & +P_{3}\sum_{R''}\eta_{R''}^{(3)}\left(\left|\langle J_{z,1}\rangle_{R''}\right|+\left|\langle J_{z,2}\rangle_{R''}\right|+2\left|\langle J_{z,3}\rangle_{R''}\right|\right)\nonumber \\
 & \geq & \left|\langle J_{z,1}\rangle\right|+\left|\langle J_{z,2}\rangle\right|+\left|\langle J_{z,3}\rangle\right|+P_{1}\sum_{R}\eta_{R}^{(1)}\left|\langle J_{z,1}\rangle_{R}\right|+P_{2}\sum_{R'}\eta_{R'}^{(2)}\left|\langle J_{z,2}\rangle_{R'}\right|+P_{3}\sum_{R''}\eta_{R''}^{(3)}\left|\langle J_{z,3}\rangle_{R''}\right|\nonumber \\
 & \geq & \left|\langle J_{z,1}\rangle\right|+\left|\langle J_{z,2}\rangle\right|+\left|\langle J_{z,3}\rangle\right|
\end{eqnarray}
where we use that $\langle J_{z,k}\rangle=P_{1}\sum_{R}\eta_{R}^{(1)}\langle J_{z,k}\rangle_{R}+P_{2}\sum_{R'}\eta_{R'}^{(2)}\langle J_{z,k}\rangle_{R'}+P_{3}\sum_{R''}\eta_{R''}^{(3)}\langle J_{z,k}\rangle_{R''}$
and hence $|\langle J_{z,k}\rangle|\leq P_{1}\sum\eta_{R}^{(1)}|\langle J_{z,k}\rangle_{R}|+P_{2}\sum_{R'}\eta_{R'}^{(2)}|\langle J_{z,k}\rangle_{R'}|+P_{3}\sum_{R''}\eta_{R''}^{(3)}|\langle J_{z,k}\rangle_{R''}|$.

It follows that the criterion is also for full tripartite inseparability,
since full tripartite inseparability is a special case of genuine
tripartite entanglement. $\square$
\end{widetext}

The number of moment measurements in the criterion given by Eq. (\ref{eq:genuine_tripartite_vlf_sums})
can be reduced by using a criterion that only involves two of the
three quantities $B_{I}$, $B_{II}$ and $B_{III}$. For the limiting
case of $\left|\langle J_{z,k}\rangle\right|\rightarrow\infty$ and
$\Delta^{2}J_{z,k}/\left|\langle J_{z,k}\rangle\right|\le1$, we can
use the inequalities (\ref{eq:vlF_sum}) to deduce that 
\begin{align}
B_{I}+B_{II} & \geq\left|\langle J_{z,1}\rangle\right|+2\left|\langle J_{z,2}\rangle\right|+\left|\langle J_{z,3}\rangle\right|\label{eq:B1+B2-1}
\end{align}
is satisfied by any mixture of all tripartite biseparable states.
The violation of the criterion in Eq. (\ref{eq:B1+B2-1}) then implies
genuine tripartite entanglement. This is also true for other combinations
$B_{I}+B_{III}\geq2\left|\langle J_{z,1}\rangle\right|+\left|\langle J_{z,2}\rangle\right|+\left|\langle J_{z,3}\rangle\right|$
and $B_{II}+B_{III}\geq\left|\langle J_{z,1}\rangle\right|+\left|\langle J_{z,2}\rangle\right|+2\left|\langle J_{z,3}\rangle\right|$.

However, as before, we can derive an inequality that is valid generally,
\emph{without} the assumption $\left|\langle J_{z,k}\rangle\right|\rightarrow\infty$
and $\Delta^{2}J_{z,k}/\left|\langle J_{z,k}\rangle\right|\le1$.
Setting $g_{1}=g_{2}=g_{3}=1$, we see from Eqs. (\ref{eq:B-1}),
(\ref{eq:B-2}) and (\ref{eq:B-3}) that 
\begin{align*}
B_{I}+B_{II} & \geq P_{1}\sum_{R}\eta_{R}^{(1)}\left[\left|\langle J_{z,1}\rangle_{R}\right|+\left|\langle J_{z,2}\rangle_{R}\right|\right]\\
 & +P_{2}\sum_{R'}\eta_{R'}^{(2)}\left[\left|\langle J_{z,1}\rangle_{R'}\right|+\left|\langle J_{z,2}\rangle\right|_{R'}\right]\\
 & +P_{2}\sum_{R'}\eta_{R'}^{(2)}\left[\left|\langle J_{z,2}\rangle_{R'}\right|+\left|\langle J_{z,3}\rangle\right|_{R'}\right]\\
 & +P_{3}\sum_{R''}\eta_{R''}^{(3)}\left[\left|\langle J_{z,2}\rangle_{R''}\right|+\left|\langle J_{z,3}\rangle\right|_{R''}\right]
\end{align*}
Continuing, we find
\begin{align}
B_{I}+B_{II} & \geq P_{1}\sum_{R}\eta_{R}^{(1)}\left|\langle J_{z,2}\rangle_{R}\right|+P_{2}\sum_{R'}\eta_{R'}^{(2)}\left|\langle J_{z,2}\rangle_{R'}\right|\nonumber \\
 & +P_{3}\sum_{R''}\eta_{R''}^{(3)}\left|\langle J_{z,2}\rangle_{R''}\right|\nonumber \\
 & \geq\left|\langle J_{z,2}\rangle\right|\label{eq:B1+B2}
\end{align}
The triangle inequality is used to obtain the last line. This inequality
is thus satisfied by any mixture of all tripartite biseparable states.
Thus, we have proved the following criterion, which holds true for
an arbitrary state.

\textbf{\emph{Criterion 2b:}} The violation of the inequality $B_{I}+B_{II}\geq\left|\langle J_{z,2}\rangle\right|$
in Eq. (\ref{eq:B1+B2}) implies full tripartite inseparability and
genuine tripartite entanglement. This is also true for other combinations
$B_{I}+B_{III}\geq\left|\langle J_{z,1}\rangle\right|$ and $B_{II}+B_{III}\geq\left|\langle J_{z,3}\rangle\right|$.

\subsection{The product inequalities}

\subsubsection{Product of two variances}

Criteria involving products rather than sums can also be derived.
Again, we consider the two quantities $\Delta^{2}u=\Delta^{2}\left(h_{1}J_{x,1}+h_{2}J_{x,2}+h_{3}J_{x,3}\right)$
and $\Delta^{2}v=\Delta^{2}\left(g_{1}J_{y,1}+g_{2}J_{y,2}+g_{3}J_{y,3}\right)$.The
product of two variances $\Delta^{2}u$ and $\Delta^{2}v$ satisfies
the inequality 
\begin{align}
\Delta^{2}u\Delta^{2}v & \geq\left[\sum_{R}P_{R}\Delta^{2}u_{R}\right]\left[\sum_{R}P_{R}\Delta^{2}v_{R}\right]\nonumber \\
 & \geq\sum_{R}P_{R}\Delta^{2}u_{R}\Delta^{2}v_{R}\,,\label{eq:varu_varv-1}
\end{align}
where the Cauchy-Schwarz inequality is used. For an arbitrary bipartition
$\rho_{k}^{\zeta}\rho_{lm}^{\zeta}$, $\Delta^{2}u_{\zeta}\Delta^{2}v_{\zeta}$
satisfies the inequality (see Appendix 2): 
\begin{align}
\Delta^{2}u_{\zeta}\Delta^{2}v_{\zeta} & \geq\!\frac{1}{4}\left[\left|g_{k}h_{k}\langle J_{z,k}\rangle_{\zeta}\right|\!+\!\left|g_{l}h_{l}\langle J_{z,l}\rangle_{\zeta}+g_{m}h_{m}\langle J_{z,m}\rangle_{\zeta}\right|\right]^{2}.\label{eq:arbitrary_bipartition-1}
\end{align}
Again, provided that $\left|\langle J_{z,k}\rangle\right|\rightarrow\infty$
and $(\Delta J_{z,k})^{2}/\left|\langle J_{z,k}\rangle\right|\le1$,
the following criterion can be obtained along similar lines to the
proof given in Section II.A above. We omit the proof and only present
the result here.

\textbf{\textit{Criterion }}\textbf{\emph{3}}\label{3}. Full tripartite
inseparability is observed if the inequality 
\begin{align}
\Delta u\Delta v & \geq\frac{1}{2}\text{min}\left\{ \left|g_{1}h_{1}\langle J_{z,1}\rangle\right|+\left|g_{2}h_{2}\langle J_{z,2}\rangle+g_{3}h_{3}\langle J_{z,3}\rangle\right|\,,\right.\nonumber \\
 & \left|g_{2}h_{2}\langle J_{z,2}\rangle\right|+\left|g_{1}h_{1}\langle J_{z,1}\rangle+g_{3}h_{3}\langle J_{z,3}\rangle\right|,\nonumber \\
 & \left.\left|g_{3}h_{3}\langle J_{z,3}\rangle\right|+\left|g_{1}h_{1}\langle J_{z,1}\rangle+g_{2}h_{2}\langle J_{z,2}\rangle\right|\right\} \,\label{eq:criterion-2}
\end{align}
is violated. Genuine tripartite entanglement is confirmed if the inequality
is violated in the limit where the mean spins are large, $\left|\langle J_{z,k}\rangle\right|\rightarrow\infty$,
and where fluctuations satisfy $\Delta^{2}J_{z,k}/\left|\langle J_{z,k}\rangle\right|\le1$.

\subsubsection{Van Loock-Furusawa product inequalities}

The product version of the van Loock-Furusawa inequalities can be
obtained, with proofs that proceed in the same way as the corresponding
van Loock-Furusawa sum inequalities. The quantities involved are $S_{I}$,
$S_{II}$, and $S_{III}$, as defined below: 
\begin{align}
S_{I} & \equiv\Delta\left(J_{x,1}-J_{x,2}\right)\Delta\left(J_{y,1}+J_{y,2}+g_{3}J_{y,3}\right)\nonumber \\
S_{II} & \equiv\Delta\left(J_{x,2}-J_{x,3}\right)\Delta\left(g_{1}J_{y,1}+J_{y,2}+J_{y,3}\right)\nonumber \\
S_{III} & \equiv\Delta\left(J_{x,1}-J_{x,3}\right)\Delta\left(J_{y,1}+g_{2}J_{y,2}+J_{y,3}\right)\,.\label{eq:vlf_S}
\end{align}
First, we look at the case with $\left|\langle J_{z,k}\rangle\right|\rightarrow\infty$
and $(\Delta J_{z,k})^{2}/\left|\langle J_{z,k}\rangle\right|\le1$.
By choosing the coefficients $g_{i}$ and $h_{i}$ in Eq. (\ref{eq:criterion-2}),
we obtain a set of inequalities satisfied by $S_{I}$, $S_{II}$ and
$S_{III}$. For example, the left side of the criterion in Eq. (\ref{eq:criterion-2})
is equal to $S_{I}$ when $h_{1}=1,\,h_{2}=-1,\,h_{3}=0$ and $g_{1}=g_{2}=1$.
Similar to Eq. (\ref{eq:vlf_inseparability}), the product version
of the set of van Loock-Furusawa inequalities $S_{I}$, $S_{II}$
and $S_{III}$ satisfy the following inequalities\textit{\textcolor{black}{:
\begin{align}
S_{I} & \geq\frac{1}{2}\left(\left|\langle J_{z,1}\rangle\right|+\left|\langle J_{z,2}\rangle\right|\right)\nonumber \\
S_{II} & \geq\frac{1}{2}\left(\left|\langle J_{z,2}\rangle\right|+\left|\langle J_{z,3}\rangle\right|\right)\nonumber \\
S_{III} & \geq\frac{1}{2}\left(\left|\langle J_{z,1}\rangle\right|+\left|\langle J_{z,3}\rangle\right|\right)\,,\label{eq:vlF_product}
\end{align}
}}and the violation of any two of these inequalities implies full
tripartite inseparability. Again, following the same derivation as
in the sum version of van Loock-Furusawa set of inequalities, when
the biseparable mixture state is taken into account, the product version
of the set of inequalities is given below:\textit{ 
\begin{align}
S_{I} & \geq\frac{1}{2}\left(P_{1}+P_{2}\right)\left(\left|\langle J_{z,1}\rangle\right|+\left|\langle J_{z,2}\rangle\right|\right)\nonumber \\
S_{II} & \geq\frac{1}{2}\left(P_{2}+P_{3}\right)\left(\left|\langle J_{z,2}\rangle\right|+\left|\langle J_{z,3}\rangle\right|\right)\nonumber \\
S_{III} & \geq\frac{1}{2}\left(P_{1}+P_{3}\right)\left(\left|\langle J_{z,1}\rangle\right|+\left|\langle J_{z,3}\rangle\right|\right)\,,\label{eq:vlF_product-1}
\end{align}
}

Genuine tripartite entanglement is observed if the inequality $S_{I}+S_{II}+S_{III}\geq\frac{1}{2}\left(\left|\langle J_{z,1}\rangle\right|+\left|\langle J_{z,2}\rangle\right|+\left|\langle J_{z,3}\rangle\right|\right)\;$
is violated. This is first proved in the limiting case $\left|\langle J_{z,k}\rangle\right|\rightarrow\infty$
and $(\Delta J_{z,k})^{2}/\left|\langle J_{z,k}\rangle\right|\le1$
below.

\emph{Proof.} Taking the sum of $S_{I}$, $S_{II}$ and $S_{III}$
in Eq. (\ref{eq:vlF_product-1}), we obtain 
\begin{align*}
S_{I}+S_{II}+S_{III} & \geq\frac{1}{2}\left(2P_{1}+P_{2}+P_{3}\right)\left|\langle J_{z,1}\rangle\right|\\
 & +\frac{1}{2}\left(P_{1}+2P_{2}+P_{3}\right)\left|\langle J_{z,2}\rangle\right|\\
 & +\frac{1}{2}\left(P_{1}+P_{2}+2P_{3}\right)\left|\langle J_{z,3}\rangle\right|
\end{align*}
Continuing we see that
\begin{align*}
S_{I}+S_{II}+S_{III} & \geq\frac{1}{2}\left(2P_{1}+P_{2}+P_{3}\right)\left|\langle J_{z,1}\rangle\right|\\
 & \ \ \ \times\left(\left|\langle J_{z,1}\rangle\right|+\left|\langle J_{z,2}\rangle\right|+\left|\langle J_{z,3}\rangle\right|\right)\\
 & =\frac{1}{2}\left(\left|\langle J_{z,1}\rangle\right|+\left|\langle J_{z,2}\rangle\right|+\left|\langle J_{z,3}\rangle\right|\right)
\end{align*}
as required. $\square$

Next, we derive the van-Loock Furusawa type product inequalities in
the general case, \emph{without} the assumption $\left|\langle J_{z,k}\rangle\right|\rightarrow\infty$
and $(\Delta J_{z,k})^{2}/\left|\langle J_{z,k}\rangle\right|\le1$.\bigskip{}

\textbf{\textit{\textcolor{black}{Criterion 4\label{4}.}}}\textit{\textcolor{black}{{}
}}Genuine tripartite entanglement and full tripartite inseparability
are observed if the following inequality is violated: 
\begin{align}
S_{I}+S_{II}+S_{III} & \geq\frac{1}{2}\left(\left|\langle J_{z,1}\rangle\right|+\left|\langle J_{z,2}\rangle\right|+\left|\langle J_{z,3}\rangle\right|\right)\,.\label{eq:genuine_tripartite_vlf_sumofproducts}
\end{align}
\textit{Proof}. We need only prove for genuine tripartite entanglement,
since full tripartite inseparability follows as a special case. By
choosing $h_{1}=g_{1}=-h_{2}=g_{2}=1$ and $h_{3}=0$, inequality
Eq. (\ref{eq:varu_varv-1}) has the expression 
\begin{align}
S_{I} & \geq\frac{1}{2}P_{1}\sum_{R}\eta_{R}^{(1)}\left[\left|\langle J_{z,1}\rangle_{R}\right|+\left|\langle J_{z,2}\rangle_{R}\right|\right]\nonumber \\
 & +\frac{1}{2}P_{2}\sum_{R'}\eta_{R'}^{(2)}\left[\left|\langle J_{z,1}\rangle_{R'}\right|+\left|\langle J_{z,2}\rangle\right|_{R'}\right]\,\label{eq:S-1}
\end{align}
Similarly choosing $h_{2}=g_{2}=g_{3}=-h_{3}=1$ and $h_{1}=0$ 
\begin{align}
S_{II} & \equiv\Delta\left(J_{x,2}-J_{x,3}\right)\Delta\left(g_{1}J_{y,1}+J_{y,2}+J_{y,3}\right)\nonumber \\
 & \geq\frac{1}{2}P_{2}\sum_{R'}\eta_{R'}^{(2)}\left[\left|\langle J_{z,2}\rangle_{R'}\right|+\left|\langle J_{z,3}\rangle\right|_{R'}\right]\nonumber \\
 & +\frac{1}{2}P_{3}\sum_{R''}\eta_{R''}^{(3)}\left[\left|\langle J_{z,2}\rangle_{R''}\right|+\left|\langle J_{z,3}\rangle\right|_{R''}\right]\label{eq:S-2}
\end{align}
Similarly, 
\begin{align}
S_{III} & \equiv\Delta\left(J_{x,1}-J_{x,3}\right)\Delta\left(J_{y,1}+g_{2}J_{y,2}+J_{y,3}\right)\nonumber \\
 & \geq\frac{1}{2}P_{1}\sum_{R}\eta_{R}^{(1)}\left[\left|\langle J_{z,1}\rangle_{R}\right|+\left|\langle J_{z,3}\rangle\right|_{R}\right]\nonumber \\
 & +\frac{1}{2}P_{3}\sum_{R''}\eta_{R''}^{(3)}\left[\left|\langle J_{z,1}\rangle_{R''}\right|+\left|\langle J_{z,3}\rangle\right|_{R''}\right]\,\label{eq:S-3}
\end{align}
To determine the criterion for genuine tripartite entanglement, we
then see that
\begin{widetext}
\begin{eqnarray}
S_{I}+S_{II}+S_{III} & \geq & \frac{1}{2}P_{1}\sum_{R}\eta_{R}^{(1)}\left(2\left|\langle J_{z,1}\rangle_{R}\right|+\left|\langle J_{z,2}\rangle_{R}\right|+\left|\langle J_{z,3}\rangle_{R}\right|\right)+\frac{1}{2}P_{2}\sum_{R'}\eta_{R'}^{(2)}\left(\left|\langle J_{z,1}\rangle_{R'}\right|+2\left|\langle J_{z,2}\rangle_{R'}\right|+\left|\langle J_{z,3}\rangle_{R'}\right|\right)\nonumber \\
 &  & +\frac{1}{2}P_{3}\sum_{R''}\eta_{R''}^{(3)}\left(\left|\langle J_{z,1}\rangle_{R''}\right|+\left|\langle J_{z,2}\rangle_{R''}\right|+2\left|\langle J_{z,3}\rangle_{R''}\right|\right)\nonumber \\
 & = & \frac{1}{2}\left(\left|\langle J_{z,1}\rangle\right|+\left|\langle J_{z,2}\rangle\right|+\left|\langle J_{z,3}\rangle\right|\right)+P_{1}\sum_{R}\eta_{R}^{(1)}\left|\langle J_{z,1}\rangle_{R}\right|+P_{2}\sum_{R'}\eta_{R'}^{(2)}\left|\langle J_{z,2}\rangle_{R'}\right|+P_{3}\sum_{R''}\eta_{R''}^{(3)}\left|\langle J_{z,3}\rangle_{R''}\right|\nonumber \\
 & \geq & \frac{1}{2}\left(\left|\langle J_{z,1}\rangle\right|+\left|\langle J_{z,2}\rangle\right|+\left|\langle J_{z,3}\rangle\right|\right)\,.\ \square\label{eq:vlf_prod}
\end{eqnarray}
\end{widetext}

The number of moment measurements in the criterion can be reduced
by using a criterion that only involves two of the three quantities
$S_{I}$, $S_{II}$ and $S_{III}$. Following the proof of Criterion
2b, we obtain the following Criterion which hold true for all states.

\textbf{\emph{Criterion 4b:}} The violation of the inequality $S_{I}+S_{II}\geq\left|\langle J_{z,2}\rangle\right|/2$
implies genuine tripartite entanglement. This is also true for other
combinations $S_{I}+S_{III}\geq\left|\langle J_{z,1}\rangle\right|/2$
and $S_{II}+S_{III}\geq\left|\langle J_{z,3}\rangle\right|/2$.

\section{Inequalities involving planar spin uncertainty relations}

The inequalities in the previous two sections used the canonical spin
uncertainty relations. For certain quantum states such as the multipartite
spin GHZ state, the right side of these inequalities might be zero,
giving the trivial relation that a sum or product of variances should
be positive. Here, we consider the planar uncertainty relation, where
the sum of uncertainties in two of the orthogonal spin observables
has a lower bound that is a function of the spin value of the state.
The planar uncertainty relation was obtained for spin $J=1/2$ \cite{PhysRevA.35.1486}
and $J=1$ \cite{Hofmann_PRA2003}, and was later calculated for an
arbitrary spin $J$ by He et al. \cite{PhysRevA.84.022107}. In that
work, they minimized $\Delta^{2}J_{x}+\Delta^{2}J_{y}$ for a general
quantum state written in the spin-$z$ basis as 
\begin{align}
|\psi\rangle & =\frac{1}{\sqrt{n}}\sum_{m=-J}^{J}R_{m}e^{-i\phi_{m}}|J,m\rangle\,,\label{eq:single_particle_state}
\end{align}
Here $R_{m},\phi_{m}$ are real numbers characterizing the amplitude
and phase of the basis state $|J,m\rangle$, while $n$ is the normalization
factor given by $n=\sum_{m=-J}^{J}R_{m}^{2}$. He et al. found the
lower bound $C_{J}$ ($C_{J}>0$) such that for a given $J$ 
\begin{equation}
\Delta^{2}J_{x}+\Delta^{2}J_{y}\geq C_{J}\label{eq:psur}
\end{equation}
Also in that work \cite{PhysRevA.84.022107}, a criterion that verifies
the $N$-partite inseparability was derived. Since the total $N$-partite
separable state is a probabilistic sum of tensor product of $N$ density
operators, the planar uncertainty relation can be used. This is not
the case for genuine multipartite entanglement where a biseparable
state contains partitions that cannot be expressed as a product state
of those particles/ modes in those partitions.

Nevertheless, the planar uncertainty relation can be used to detect
genuine tripartite entanglement, if we use an inference variance method
\cite{Reid_PRA1989,Reid_colloqium_RMP2009}.

\textbf{\emph{Criterion 5.}} \textcolor{black}{Consider the inequality
given by 
\begin{align}
B_{1}+B_{2}+B_{3} & \geq C_{J}\,,\label{eq:tripartite_planar-1-1}
\end{align}
where 
\begin{eqnarray*}
B_{1} & = & \Delta^{2}\left(J_{x,1}-O_{23}^{(1)}\right)+\Delta^{2}\left(J_{y,1}-P_{23}^{(1)}\right)\\
B_{2} & = & \Delta^{2}\left(J_{x,2}-O_{13}^{(2)}\right)+\Delta^{2}\left(J_{y,2}-P_{13}^{(2)}\right)\\
B_{3} & = & \Delta^{2}\left(J_{x,3}-O_{12}^{(3)}\right)+\Delta^{2}\left(J_{y,3}-P_{12}^{(3)}\right)
\end{eqnarray*}
and $O_{lm}^{(k)}$, $P_{lm}^{(k)}$ are observables defining measurements
that can be made on the combined subsystems that we denote by $l$
and $m$. The violation of this inequality suffices to confirm genuine
tripartite entanglement of the three systems denoted $1$, $2$ and
$3$. }\textit{\textcolor{black}{\emph{Full tripartite inseparability
is observed if 
\begin{equation}
B_{k}\geq C_{J}\label{eq:fis}
\end{equation}
for each $k=1,2,3$.}}}

\textit{\textcolor{black}{Proof.}}\textcolor{black}{{} Consider $\Delta^{2}\left(J_{x,1}-O_{23}^{(1)}\right)$
and $\Delta^{2}\left(J_{y,1}-P_{23}^{(1)}\right)$ where $O_{23}^{(1)}$
and $P_{23}^{(1)}$ are operators for systems $2$ and $3$. We derive
the following inequality that holds for an arbitrary pure state with
a separable bipartition $\rho_{1}^{\zeta}\rho_{23}^{\zeta}$. 
\begin{align}
B_{1} & =\Delta^{2}\left(J_{x,1}-O_{23}^{(1)}\right)+\Delta^{2}\left(J_{y,1}-P_{23}^{(1)}\right)\nonumber \\
 & \geq\Delta^{2}\left(J_{x,1}\right)+\Delta^{2}\left(J_{y,1}\right)\nonumber \\
 & \geq C_{J}\label{eq:B_1}
\end{align}
This holds also for all mixtures of separable bipartitions $\rho_{1}^{\zeta}\rho_{23}^{\zeta}$.
Similarly, the inequalities 
\begin{equation}
B_{2}\geq\Delta^{2}\left(J_{x,2}-O_{13}^{(2)}\right)+\Delta^{2}\left(J_{y,2}-P_{13}^{(2)}\right)\geq C_{J}\label{eq:B_2}
\end{equation}
and 
\begin{equation}
B_{3}\geq\Delta^{2}\left(J_{x,3}-O_{12}^{(3)}\right)+\Delta^{2}\left(J_{y,3}-P_{12}^{(3)}\right)\geq C_{J}\label{eq:B_3}
\end{equation}
follow from the separable bipartitions $\rho_{2}^{\zeta}\rho_{13}^{\zeta}$
and $\rho_{3}^{\zeta}\rho_{12}^{\zeta}$ respectively. For a pure
state, if all three inequalities are violated, we can conclude that
the three systems are genuinely tripartite entangled. For a mixed
state the conditions change. We require to falsify an arbitrary biseparable
mixed state given by $\rho_{BS}=P_{1}\sum_{R}\eta_{R}^{\left(1\right)}\rho_{1}^{R}\rho_{23}^{R}+P_{2}\sum_{R'}\eta_{R'}^{\left(2\right)}\rho_{2}^{R'}\rho_{13}^{R'}+P_{3}\sum_{R''}\eta_{R''}^{(3)}\rho_{3}^{R''}\rho_{12}^{R''}$,
as defined by Eq. (\ref{eq:biseparable_den_op}). We give a proof
similar to those given for Criteria 2 and 4. For brevity, we index
the biseparable states $\sum_{R}\eta_{R}^{\left(1\right)}\rho_{1}^{R}\rho_{23}^{R}$,
$\sum_{R'}\eta_{R'}^{\left(2\right)}\rho_{2}^{R'}\rho_{13}^{R'}$
and $\sum_{R''}\eta_{R''}^{(3)}\rho_{3}^{R''}\rho_{12}^{R''}$ by
$k=1,2,3$, respectively. Thus, we denote $B_{1,1}$ to be the quantity
$B_{1}$ that is evaluated using the biseparable state $\sum_{R}\eta_{R}^{\left(1\right)}\rho_{1}^{R}\rho_{23}^{R}$.
Then, for the biseparable mixture, 
\begin{align*}
B_{1} & \geq\sum_{k}P_{k}B_{1,k}\\
 & \geq P_{1}B_{1,1}\geq P_{1}C_{J}\,.
\end{align*}
Similarly, for a biseparable mixture, $B_{2}\geq P_{2}C_{J}$ and
$B_{3}\geq P_{3}C_{J}$. In order to include all possible biseparable
mixtures, we consider 
\begin{align*}
B_{1}+B_{2}+B_{3} & \geq\left(P_{1}+P_{2}+P_{3}\right)C_{J}=C_{J}
\end{align*}
using $\sum_{k}P_{k}=1$. Thus, all biseparable mixtures are excluded
when this inequality is violated. $\square$}

\textcolor{black}{This inequality has been derived in Ref. \cite{He_Reid_steering_PRL2013}
in a similar context, to give a condition for genuine tripartite steering.
Steering is a form of entanglement linked to the Einstein-Podolsky-Rosen
paradox, and hence a steering criterion will also be a criterion for
entanglement \cite{Wiseman_PRL2007}. The entanglement criterion might
be made stronger, if one can make use of uncertainty relations for
the operators $O_{lm}^{(k)}$ and $P_{lm}^{(k)}$ once these are established
for a given scenario.}

\textcolor{black}{It is straightforward to see that the inequality
is violated for the GHZ state \cite{Greenberger_AJP1990}, }\textcolor{red}{{} }\textcolor{black}{defined
as 
\begin{equation}
{\color{black}|\psi\rangle=\frac{1}{\sqrt{2}}\left(|\uparrow\uparrow\uparrow\rangle-|\downarrow\downarrow\downarrow\rangle\right)}\label{eq:ghz}
\end{equation}
where $|\uparrow\uparrow\uparrow\rangle$ ($|\downarrow\downarrow\downarrow\rangle$)
is the state with $z$-spins up (down) for all subsystems $k=1,2,3$.
This is because, as is well-known for the GHZ state, the $z$-spin,
$x$-spin and the $y$-spin of any of the three subsystems can be
inferred by joint measurements made on the other two subsystems. This
result is clear for inferring the value of $J_{z,k}$. The inequality
(\ref{eq:tripartite_planar-1-1}) applies for all spin pairs, and
if we replace $J_{y,i}$ with $J_{z,i}$, it is clear that by taking
$P_{lm}^{(k)}=J_{z}^{(l)}$, one can achieve $\Delta^{2}\left(J_{z,k}-P_{lm}^{(k)}\right)=0$
for each $k$. For inferring $J_{x,k}$, it is also clear, since the
GHZ state is an eigenstate of $J_{x,1}J_{x,2}J_{x,3}$ with eigenvalue
$-1$. Thus, $O_{lm}^{(k)}$ is the measurement given as follows:
Measure the spin $J_{x}$ of each of the other subsystems $l$ and
$m$, and assign the value of the measurement by multiplying the spins
values together. If the product is $+1$, then the outcome of $O_{lm}^{(k)}$
is $-1$. If the product is $-1$, then the outcome of $O_{lm}^{(k)}$
is $+1$. In this way, we see that $\Delta^{2}\left(J_{x,k}-O_{lm}^{(k)}\right)=0$,
for each $k=1,2,3$ with $l\neq m\neq k$. }Hence, the inequality
\textcolor{black}{(\ref{eq:tripartite_planar-1-1})} is violated,
giving a simple method to detect the genuine tripartite entanglement
of GHZ states (or approximate GHZ states) in an experiment.

\textcolor{black}{We may ask whether the inequality is also violated
for the $W$ state \cite{Dur_PRA2000} given by 
\begin{equation}
|W\rangle=\frac{1}{\sqrt{3}}(|\uparrow\downarrow\downarrow\rangle+|\downarrow\uparrow\downarrow\rangle+|\downarrow\downarrow\uparrow\rangle)\,.\label{eq:w-2}
\end{equation}
Here we will use the criterion expressed in Pauli spins, so that $B_{i}=\Delta^{2}\left(\sigma_{z,i}-O_{jk}^{(1)}\right)+\Delta^{2}\left(\sigma_{x,i}-P_{jk}^{(1)}\right)$
where $i\neq j\neq k$. The conditions then utilize $C_{J}=1$ since
$J=1/2$ }\cite{PhysRevA.35.1486}\textcolor{black}{. The spin $\sigma_{z}$
of system $1$ can be inferred by measuring the spin product of $2$
and $3$. We find that $\Delta^{2}(\sigma_{z,1}-O_{23}^{(1)})=0$.
Now consider that the spins $\sigma_{x}$ of systems $2$ and $3$
are simultaneously measured. We consider the measurement $P_{23}^{(1)}$
to have an outcome of $1$ if both spins are measured as $+1$; an
outcome $-1$ if the spins are measured as $-1$; and zero otherwise.
Simple calculation tells us that $\Delta^{2}(\sigma_{x,1}-P_{23}^{(1)})=\frac{1}{2}$.
By symmetry of the W state, this result holds for all permutations
of the subsystems. Thus we see that we are able to confirm entanglement
across each bipartition, since the condition (\ref{eq:B_1}) for Pauli
spins reduces to $B_{1}\geq1$. Since we find $B_{1}=B_{2}=B_{3}=\frac{1}{2}$,
the condition for tripartite inseparability is satisfied. If in an
experiment we are able to verify a pure state, then this implies genuine
tripartite entanglement. We note the above condition for mixed states,
$B_{1}+B_{2}+B_{3}<1$ is not satisfied. The $W$ state (\ref{eq:w-2})
}\textcolor{black}{\emph{is}}\textcolor{black}{{} genuinely tripartite
entangled. That the condition is not satisfied merely reflects that
the criteria we derive are sufficient, but not necessary, to certify
genuine tripartite entanglement.} \vskip -0.01in Svetlichny derived
conditions to detect the genuine tripartite entanglement of three
spin $1/2$ systems in the form of Bell inequalities. Further criteria
for the certification of the genuine tripartite entanglement of GHZ,
W and cluster states have been derived\textit{\textcolor{black}{\emph{
in Refs. \cite{Toth_PRA2005,Korbicz_PRA2006,Sperling_PRL2013}. The
method given above is not necessarily advantageous over these earlier
methods. It can be readily extended (by applying uncertainty relation
(\ref{eq:psur})) however to conditions for higher $J$.}}}

\section{Criteria for genuine $N$-partite entanglement \label{sec:N_partite}}

The method used in Section II to derive criteria for full $N$-partite
inseparability and genuine tripartite entanglement can be extended
to $N$-partite systems. The complication arises in that the set of
possible bipartitions scales as $\left(2^{N-1}-1\right)$, and every
bipartition has to be taken into account in the derivation of these
criteria that certify genuine $N$-partite entanglement.

Here, we generalize the criterion in Eq. (\ref{eq:criterion-1}) for
$N$-partite spin systems. First, we present criteria that are valid
when $\left|\langle J_{z,k}\rangle\right|\rightarrow\infty$ and $\Delta^{2}J_{z,k}/\left|\langle J_{z,k}\rangle\right|\le1$.
These criteria are Criteria 6, 7, 8, and 9. Then we present Criterion
10, which is a criterion for genuine $4$-partite entanglement without
the extra assumptions. The derivations can be readily extended to
a larger number of systems $N$.

\textbf{\textit{Criterion }}\textbf{\emph{6}}. We denote each bipartition
by $S_{r}-S_{s}$, where $S_{r}$ and $S_{s}$ are two sets of modes
in the partitions in a specific bipartition. Then, the violation of
the inequality 
\begin{align}
\Delta^{2}u+\Delta^{2}v & \geq\text{min}\left\{ S_{B}\right\} \,\label{eq:criterion_Nparitite_sum}
\end{align}
implies full $N$-partite inseparability, where $S_{B}$ is $\left(\left|\sum_{k_{r}=1}^{m}h_{k_{r}}g_{k_{r}}\langle J_{z,k_{r}}\rangle\right|+\left|\sum_{k_{s}=1}^{n}h_{k_{s}}g_{k_{s}}\langle J_{z,k_{s}}\rangle\right|\right)$.
In the limit of $\left|\langle J_{z,k}\rangle\right|\rightarrow\infty$
and $\Delta^{2}J_{z,k}/\left|\langle J_{z,k}\rangle\right|\le1$,
the violation implies genuine $N$-partite entanglement. The proof
for this inequality follows similarly to the proof for the inequality
in Eq. (\ref{eq:criterion-1}).

\textbf{\textit{Criterion }}\textbf{\emph{7}}. Similarly, the Criterion
6 holds, with the sum inequality replaced by the corresponding product
inequality given by 
\begin{align}
\Delta u\Delta v & \geq\frac{1}{2}\text{min}\left\{ S_{B}\right\} \,.\label{eq:criterion_Nparitite-product}
\end{align}

\subsection{Criteria for genuine four-partite entanglement}

\subsubsection{Sum and product inequalities}

\textbf{\textit{Criterion }}\textbf{\emph{8}}. For $N=4$, there will
be $2^{4-1}-1=7$ bipartitions. They are, using the $S_{r}-S_{s}$
notation, $1-234$, $2-134$, $3-124$, $4-123$, $12-34$, $13-24$
and $14-23$. The Criterion 6 becomes this Criterion, where the sum
inequality given by Eq. (\ref{eq:criterion_Nparitite_sum}) is
\begin{widetext}
\begin{align}
\Delta^{2}u+\Delta^{2}v & \geq\text{min}\left\{ \left|g_{1}h_{1}\langle J_{z,1}\rangle\right|+\left|g_{2}h_{2}\langle J_{z,2}\rangle+g_{3}h_{3}\langle J_{z,3}\rangle+g_{4}h_{4}\langle J_{z,4}\rangle\right|,\right.\nonumber \\
 & \left|g_{2}h_{2}\langle J_{z,2}\rangle\right|+\left|g_{1}h_{1}\langle J_{z,1}\rangle+g_{3}h_{3}\langle J_{z,3}\rangle+g_{4}h_{4}\langle J_{z,4}\rangle\right|,\nonumber \\
 & \left|g_{3}h_{3}\langle J_{z,3}\rangle\right|+\left|g_{1}h_{1}\langle J_{z,1}\rangle+g_{2}h_{2}\langle J_{z,2}\rangle+g_{4}h_{4}\langle J_{z,4}\rangle\right|,\nonumber \\
 & \left|g_{4}h_{4}\langle J_{z,4}\rangle\right|+\left|g_{1}h_{1}\langle J_{z,1}\rangle+g_{2}h_{2}\langle J_{z,2}\rangle+g_{3}h_{3}\langle J_{z,3}\rangle\right|,\nonumber \\
 & \left|g_{1}h_{1}\langle J_{z,1}\rangle+g_{2}h_{2}\langle J_{z,2}\rangle\right|+\left|g_{3}h_{3}\langle J_{z,3}\rangle+g_{4}h_{4}\langle J_{z,4}\rangle\right|,\nonumber \\
 & \left|g_{1}h_{1}\langle J_{z,1}\rangle+g_{3}h_{3}\langle J_{z,3}\rangle\right|+\left|g_{2}h_{2}\langle J_{z,2}\rangle+g_{4}h_{4}\langle J_{z,4}\rangle\right|,\nonumber \\
 & \left.\left|g_{1}h_{1}\langle J_{z,1}\rangle+g_{4}h_{4}\langle J_{z,4}\rangle\right|+\left|g_{2}h_{2}\langle J_{z,2}\rangle+g_{3}h_{3}\langle J_{z,3}\rangle\right|\right\} \equiv\text{min}\left\{ S_{B,4}\right\} \,.\label{eq:4_partite_sum}
\end{align}
\textbf{\textit{Criterion }}\textbf{\emph{9}}. Similarly, the product
inequality of Criterion 7 is given by 
\begin{align}
\Delta u\Delta v & \geq\frac{1}{2}\text{min}\left\{ S_{B,4}\right\} \,\label{eq:4_partite_product}
\end{align}
where $S_{B,4}$ is defined in Eq. (\ref{eq:4_partite_sum}). The
violation of inequality in Eq. (\ref{eq:4_partite_sum}) or Eq. (\ref{eq:4_partite_product})
implies the presence of full $4$-partite inseparability, and genuine
four-partite entanglement in the limit given by Criterion 6 and 7.
 
\end{widetext}

\subsubsection{Criteria involving van Loock-Furusawa inequalities}

 Next, we derive the van Loock-Furusawa type inequality for $N$-partite
without the assumptions of large mean spins and small fluctuations.
Van Loock and Furusawa \cite{vanLoock_PRA2003} derived a set of six
inequalities to rule out four-partite inseparability. In the limiting
case where $\left|\langle J_{z,k}\rangle\right|\rightarrow\infty$
and $\Delta^{2}J_{z,k}/\left|\langle J_{z,k}\rangle\right|\le1$,
we can show that 
\begin{widetext}
\begin{align}
B_{I} & \equiv\Delta^{2}\left(J_{x,1}-J_{x,2}\right)+\Delta^{2}\left(J_{y,1}+J_{y,2}+g_{3}J_{y,3}+g_{4}J_{y,4}\right)\geq\left(\left|\langle J_{z,1}\rangle\right|+\left|\langle J_{z,2}\rangle\right|\right)\,\nonumber \\
B_{II} & \equiv\Delta^{2}\left(J_{x,2}-J_{x,3}\right)+\Delta^{2}\left(g_{1}J_{y,1}+J_{y,2}+J_{y,3}+g_{4}J_{y,4}\right)\geq\left(\left|\langle J_{z,2}\rangle\right|+\left|\langle J_{z,3}\rangle\right|\right)\,\nonumber \\
B_{III} & \equiv\Delta^{2}\left(J_{x,1}-J_{x,3}\right)+\Delta^{2}\left(J_{y,1}+g_{2}J_{y,2}+J_{y,3}+g_{4}J_{y,4}\right)\geq\left(\left|\langle J_{z,1}\rangle\right|+\left|\langle J_{z,3}\rangle\right|\right)\,\nonumber \\
B_{IV} & \equiv\Delta^{2}\left(J_{x,3}-J_{x,4}\right)+\Delta^{2}\left(g_{1}J_{y,1}+g_{2}J_{y,2}+J_{y,3}+J_{y,4}\right)\geq\left(\left|\langle J_{z,3}\rangle\right|+\left|\langle J_{z,4}\rangle\right|\right)\,\nonumber \\
B_{V} & \equiv\Delta^{2}\left(J_{x,2}-J_{x,4}\right)+\Delta^{2}\left(g_{1}J_{y,1}+J_{y,2}+g_{3}J_{y,3}+J_{y,4}\right)\geq\left(\left|\langle J_{z,2}\rangle\right|+\left|\langle J_{z,4}\rangle\right|\right)\,\nonumber \\
B_{VI} & \equiv\Delta^{2}\left(J_{x,1}-J_{x,4}\right)+\Delta^{2}\left(J_{y,1}+g_{2}J_{y,2}+g_{3}J_{y,3}+J_{y,4}\right)\geq\left(\left|\langle J_{z,1}\rangle\right|+\left|\langle J_{z,4}\rangle\right|\right)\,.\label{eq:vlf_4_partite-1}
\end{align}
\end{widetext}

Following the same proof for Eq. (\ref{eq:vlf_inseparability}) for
the tripartite case, $B_{I}$ is implied by biseparable states $\rho_{1}\rho_{234}$,
$\rho_{2}\rho_{134}$, $\rho_{13}\rho_{24}$ and $\rho_{14}\rho_{23}$,
$B_{II}$ is implied by $\rho_{2}\rho_{134}$, $\rho_{3}\rho_{124}$,
$\rho_{12}\rho_{34}$ and $\rho_{24}\rho_{13}$, $B_{III}$ is implied
by $\rho_{1}\rho_{234}$, $\rho_{3}\rho_{124}$, $\rho_{12}\rho_{34}$
and $\rho_{14}\rho_{23}$, $B_{IV}$ is implied by $\rho_{3}\rho_{124}$,
$\rho_{4}\rho_{123}$, $\rho_{13}\rho_{24}$ and $\rho_{23}\rho_{14}$,
$B_{V}$ is implied by $\rho_{2}\rho_{134}$, $\rho_{4}\rho_{123}$,
$\rho_{12}\rho_{34}$ and $\rho_{23}\rho_{14}$, and $B_{VI}$ is
implied by $\rho_{1}\rho_{234}$, $\rho_{4}\rho_{123}$, $\rho_{12}\rho_{34}$
and $\rho_{13}\rho_{24}$. We note that the violation of any three
of these inequalities implies that the system cannot be in any of
the biseparable state, thus impying full 4-partite inseparability.

We can derive similar inequalities to certify genuine four-partite
entanglement. The six spin inequalities are given by 
\begin{widetext}
\begin{align*}
B_{I} & \equiv\Delta^{2}\left(J_{x,1}-J_{x,2}\right)+\Delta^{2}\left(J_{y,1}+J_{y,2}+g_{3}J_{y,3}+g_{4}J_{y,4}\right)\\
 & \geq P_{1}\left(\left|\langle J_{z,1}\rangle_{\rho_{I}}\right|+\left|\langle J_{z,2}\rangle_{\rho_{I}}\right|\right)+P_{2}\left(\left|\langle J_{z,1}\rangle_{\rho_{II}}\right|+\left|\langle J_{z,2}\rangle_{\rho_{II}}\right|\right)\\
 & \quad+P_{6}\left(\left|\langle J_{z,1}\rangle_{\rho_{VI}}\right|+\left|\langle J_{z,2}\rangle_{\rho_{VI}}\right|\right)+P_{7}\left(\left|\langle J_{z,1}\rangle_{\rho_{VII}}\right|+\left|\langle J_{z,2}\rangle_{\rho_{VII}}\right|\right)\\
\\
B_{II} & \equiv\Delta^{2}\left(J_{x,2}-J_{x,3}\right)+\Delta^{2}\left(g_{1}J_{y,1}+J_{y,2}+J_{y,3}+g_{4}J_{y,4}\right)\\
 & \geq P_{2}\left(\left|\langle J_{z,2}\rangle_{\rho_{II}}\right|+\left|\langle J_{z,3}\rangle_{\rho_{II}}\right|\right)+P_{3}\left(\left|\langle J_{z,2}\rangle_{\rho_{III}}\right|+\left|\langle J_{z,3}\rangle_{\rho_{III}}\right|\right)\\
 & \quad+P_{5}\left(\left|\langle J_{z,2}\rangle_{\rho_{V}}\right|+\left|\langle J_{z,3}\rangle_{\rho_{V}}\right|\right)+P_{6}\left(\left|\langle J_{z,2}\rangle_{\rho_{VI}}\right|+\left|\langle J_{z,3}\rangle_{\rho_{VI}}\right|\right)\\
\\
B_{III} & \equiv\Delta^{2}\left(J_{x,1}-J_{x,3}\right)+\Delta^{2}\left(J_{y,1}+g_{2}J_{y,2}+J_{y,3}+g_{4}J_{y,4}\right)\\
 & \geq P_{1}\left(\left|\langle J_{z,1}\rangle_{\rho_{I}}\right|+\left|\langle J_{z,3}\rangle_{\rho_{I}}\right|\right)+P_{3}\left(\left|\langle J_{z,1}\rangle_{\rho_{III}}\right|+\left|\langle J_{z,3}\rangle_{\rho_{III}}\right|\right)\\
 & \quad+P_{5}\left(\left|\langle J_{z,1}\rangle_{\rho_{V}}\right|+\left|\langle J_{z,3}\rangle_{\rho_{V}}\right|\right)+P_{7}\left(\left|\langle J_{z,1}\rangle_{\rho_{VII}}\right|+\left|\langle J_{z,3}\rangle_{\rho_{VII}}\right|\right)
\end{align*}
\begin{align}
B_{IV} & \equiv\Delta^{2}\left(J_{x,3}-J_{x,4}\right)+\Delta^{2}\left(g_{1}J_{y,1}+g_{2}J_{y,2}+J_{y,3}+J_{y,4}\right)\nonumber \\
 & \geq P_{3}\left(\left|\langle J_{z,3}\rangle_{\rho_{III}}\right|+\left|\langle J_{z,4}\rangle_{\rho_{III}}\right|\right)+P_{4}\left(\left|\langle J_{z,3}\rangle_{\rho_{IV}}\right|+\left|\langle J_{z,4}\rangle_{\rho_{IV}}\right|\right)\nonumber \\
 & \quad+P_{6}\left(\left|\langle J_{z,3}\rangle_{\rho_{VI}}\right|+\left|\langle J_{z,4}\rangle_{\rho_{VI}}\right|\right)+P_{7}\left(\left|\langle J_{z,3}\rangle_{\rho_{VII}}\right|+\left|\langle J_{z,4}\rangle_{\rho_{VII}}\right|\right)\nonumber \\
\nonumber \\
B_{V} & \equiv\Delta^{2}\left(J_{x,2}-J_{x,4}\right)+\Delta^{2}\left(g_{1}J_{y,1}+J_{y,2}+g_{3}J_{y,3}+J_{y,4}\right)\nonumber \\
 & \geq P_{2}\left(\left|\langle J_{z,2}\rangle_{\rho_{II}}\right|+\left|\langle J_{z,4}\rangle_{\rho_{II}}\right|\right)+P_{4}\left(\left|\langle J_{z,2}\rangle_{\rho_{IV}}\right|+\left|\langle J_{z,4}\rangle_{\rho_{IV}}\right|\right)\nonumber \\
 & \quad+P_{5}\left(\left|\langle J_{z,2}\rangle_{\rho_{V}}\right|+\left|\langle J_{z,4}\rangle_{\rho_{V}}\right|\right)+P_{7}\left(\left|\langle J_{z,2}\rangle_{\rho_{VII}}\right|+\left|\langle J_{z,4}\rangle_{\rho_{VII}}\right|\right)\nonumber \\
\nonumber \\
B_{VI} & \equiv\Delta^{2}\left(J_{x,1}-J_{x,4}\right)+\Delta^{2}\left(J_{y,1}+g_{2}J_{y,2}+g_{3}J_{y,3}+J_{y,4}\right)\nonumber \\
 & \geq P_{1}\left(\left|\langle J_{z,1}\rangle_{\rho_{I}}\right|+\left|\langle J_{z,4}\rangle_{\rho_{I}}\right|\right)+P_{4}\left(\left|\langle J_{z,1}\rangle_{\rho_{IV}}\right|+\left|\langle J_{z,4}\rangle_{\rho_{IV}}\right|\right)\nonumber \\
 & \quad+P_{5}\left(\left|\langle J_{z,1}\rangle_{\rho_{V}}\right|+\left|\langle J_{z,4}\rangle_{\rho_{V}}\right|\right)+P_{6}\left(\left|\langle J_{z,1}\rangle_{\rho_{VI}}\right|+\left|\langle J_{z,4}\rangle_{\rho_{VI}}\right|\right)\,.\label{eq:vlf_4_partite}
\end{align}
\end{widetext}

Here, for brevity, we denote the biseparable states $\rho_{1}\rho_{234}$,
$\rho_{2}\rho_{134}$, $\rho_{3}\rho_{124}$, $\rho_{4}\rho_{123}$,
$\rho_{12}\rho_{34}$, $\rho_{13}\rho_{24}$ and $\rho_{14}\rho_{23}$
by $\rho_{I}$, $\rho_{II}$, $\rho_{III}$, $\rho_{IV}$, $\rho_{V}$,
$\rho_{VI}$, and $\rho_{VII}$ respectively. In the limit of $\left|\langle J_{z,k}\rangle\right|\rightarrow\infty$
and $\Delta^{2}J_{z,k}/\left|\langle J_{z,k}\rangle\right|\le1$,
and using the derivation as in Eq. (\ref{eq:large_spin}), Eq. (\ref{eq:vlf_4_partite})
is then given by \begin{widetext}
\begin{align}
B_{I} & \geq\left(P_{1}+P_{2}+P_{6}+P_{7}\right)\left(\left|\langle J_{z,1}\rangle\right|+\left|\langle J_{z,2}\rangle\right|\right)\,\nonumber \\
B_{II} & \geq\left(P_{2}+P_{3}+P_{5}+P_{6}\right)\left(\left|\langle J_{z,2}\rangle\right|+\left|\langle J_{z,3}\rangle\right|\right)\,\nonumber \\
B_{III} & \geq\left(P_{1}+P_{3}+P_{5}+P_{7}\right)\left(\left|\langle J_{z,1}\rangle\right|+\left|\langle J_{z,3}\rangle\right|\right)\,\nonumber \\
B_{IV} & \geq\left(P_{3}+P_{4}+P_{6}+P_{7}\right)\left(\left|\langle J_{z,3}\rangle\right|+\left|\langle J_{z,4}\rangle\right|\right)\,\nonumber \\
B_{V} & \geq\left(P_{2}+P_{4}+P_{5}+P_{7}\right)\left(\left|\langle J_{z,2}\rangle\right|+\left|\langle J_{z,4}\rangle\right|\right)\,\nonumber \\
B_{VI} & \geq\left(P_{1}+P_{4}+P_{5}+P_{6}\right)\left(\left|\langle J_{z,1}\rangle\right|+\left|\langle J_{z,4}\rangle\right|\right)\,.\label{eq:vlf_4_partite-1-1}
\end{align}
\end{widetext}We then see that the violation of the inequality 
\begin{equation}
\sum_{J=1}^{6}B_{J}\geq\left|\langle J_{z,1}\rangle\right|+\left|\langle J_{z,2}\rangle\right|+\left|\langle J_{z,3}\rangle\right|+\left|\langle J_{z,4}\rangle\right|\label{eq:sumB_J}
\end{equation}
 implies genuine 4-partite entanglement.

\emph{Proof. }Taking the sum of the set of inequalities in Eq. (\ref{eq:vlf_4_partite-1-1}),
we obtain

\begin{widetext}
\begin{align}
\sum_{J=1}^{6}B_{J} & \geq\left(P_{1}+P_{2}+P_{3}+P_{4}+P_{5}+P_{6}+P_{7}\right)\left(\left|\langle J_{z,1}\rangle\right|+\left|\langle J_{z,2}\rangle\right|+\left|\langle J_{z,3}\rangle\right|+\left|\langle J_{z,4}\rangle\right|\right)\nonumber \\
 & +\left(2P_{1}+P_{5}+P_{6}+P_{7}\right)\left|\langle J_{z,1}\rangle\right|+\left(2P_{2}+P_{5}+P_{6}+P_{7}\right)\left|\langle J_{z,2}\rangle\right|\nonumber \\
 & +\left(2P_{3}+P_{5}+P_{6}+P_{7}\right)\left|\langle J_{z,3}\rangle\right|+\left(2P_{4}+P_{5}+P_{6}+P_{7}\right)\left|\langle J_{z,4}\rangle\right|\nonumber \\
 & \geq\left(P_{1}+P_{2}+P_{3}+P_{4}+P_{5}+P_{6}+P_{7}\right)\left(\left|\langle J_{z,1}\rangle\right|+\left|\langle J_{z,2}\rangle\right|+\left|\langle J_{z,3}\rangle\right|+\left|\langle J_{z,4}\rangle\right|\right)\nonumber \\
 & =\left|\langle J_{z,1}\rangle\right|+\left|\langle J_{z,2}\rangle\right|+\left|\langle J_{z,3}\rangle\right|+\left|\langle J_{z,4}\rangle\right|\,,\label{eq:B_J_proof-1}
\end{align}
\end{widetext}where $\sum_{k}P_{k}=1$. $\square$

In fact, the criterion for genuine 4-partite entanglement can be proved
for the general case, \emph{without} the limiting assumption, as below.

\textbf{\emph{Criterion 10}}\textbf{\label{10}. }Genuine four-partite
entanglement is verified if the inequality 
\begin{align}
\sum_{J=1}^{6}B_{J} & \geq\left|\langle J_{z,1}\rangle\right|+\left|\langle J_{z,2}\rangle\right|+\left|\langle J_{z,3}\rangle\right|+\left|\langle J_{z,4}\rangle\right|\label{eq:B_J}
\end{align}
is violated. These criteria are sufficient but not necessary conditions
for four-partite inseparability, or genuine four-partite entanglement.

\emph{Proof}. From the bounds given in Eq. (\ref{eq:vlf_4_partite}),
we get 
\begin{align*}
\sum_{J=1}^{6}B_{J} & \geq\left|\langle J_{z,1}\rangle\right|+\left|\langle J_{z,2}\rangle\right|+\left|\langle J_{z,3}\rangle\right|+\left|\langle J_{z,4}\rangle\right|\,,
\end{align*}
where $\langle J_{z,k}\rangle=P_{1}\langle J_{z,k}\rangle_{\rho_{I}}+P_{2}\langle J_{z,k}\rangle_{\rho_{II}}+P_{3}\langle J_{z,k}\rangle_{\rho_{III}}+P_{4}\langle J_{z,k}\rangle_{\rho_{IV}}+P_{5}\langle J_{z,k}\rangle_{\rho_{V}}+P_{6}\langle J_{z,k}\rangle_{\rho_{VI}}+P_{7}\langle J_{z,k}\rangle_{\rho_{VII}}$.
$\square$

\section{Applications \label{sec:Applications}}

We now show how one may create $N$-partite entangled states satisfying
the criteria derived in Sections II and IV of this paper. In Section
\ref{subsec:Polarization-entanglement}, we outline optical experiments
involving polarization entanglement, where the measured observables
at each site are the Stokes operators for two polarization modes.
We then consider, in Section \ref{subsec:atomic_ensembles}, experiments
that entangle spatially-separated atomic ensembles. In Section \ref{subsec:bec_cloud},
we analyze recent experiments that generate entanglement between spatially-separated
clouds of atoms formed from a spin-squeezed Bose-Einstein condensate.
Here, for each separated subsystem, the measured observable is a Schwinger
operator involving two internal atomic levels. The Schwinger and Stokes
operators satisfy the same commutation relation as spin operators,
and hence all the criteria derived in Sections II-IV are applicable.

\subsection{\textcolor{black}{\normalsize{}{}Polarization entanglement \label{subsec:Polarization-entanglement}}}

\textcolor{black}{The polarization of a quantum state can be characterized
by the Stokes operators defined as \cite{Bowen_polarization_entanglement_PRL2002}
\begin{align}
S_{x} & =a_{H}^{\dagger}a_{H}-a_{V}^{\dagger}a_{V}\nonumber \\
S_{y} & =a_{H}^{\dagger}a_{V}e^{i\theta}+a_{V}^{\dagger}a_{H}e^{-i\theta}\nonumber \\
S_{z} & =ia_{V}^{\dagger}a_{H}e^{-i\theta}-ia_{H}^{\dagger}a_{V}e^{i\theta}\,,\label{eq:stokes_op}
\end{align}
where $a_{H}$ and $a_{V}$ are the annihilation operators of the
horizontal and vertical polarization modes respectively, and $\theta$
is the phase difference between these polarization modes. In the work
of Bowen et al. \cite{Bowen_polarization_entanglement_PRL2002}, bipartite
polarization entanglement was created by first generating CV bipartite
entanglement in the quadrature degree of freedom, and then transferring
the entanglement into the polarization degree of freedom. } 
\begin{figure}
\includegraphics[width=0.7\columnwidth]{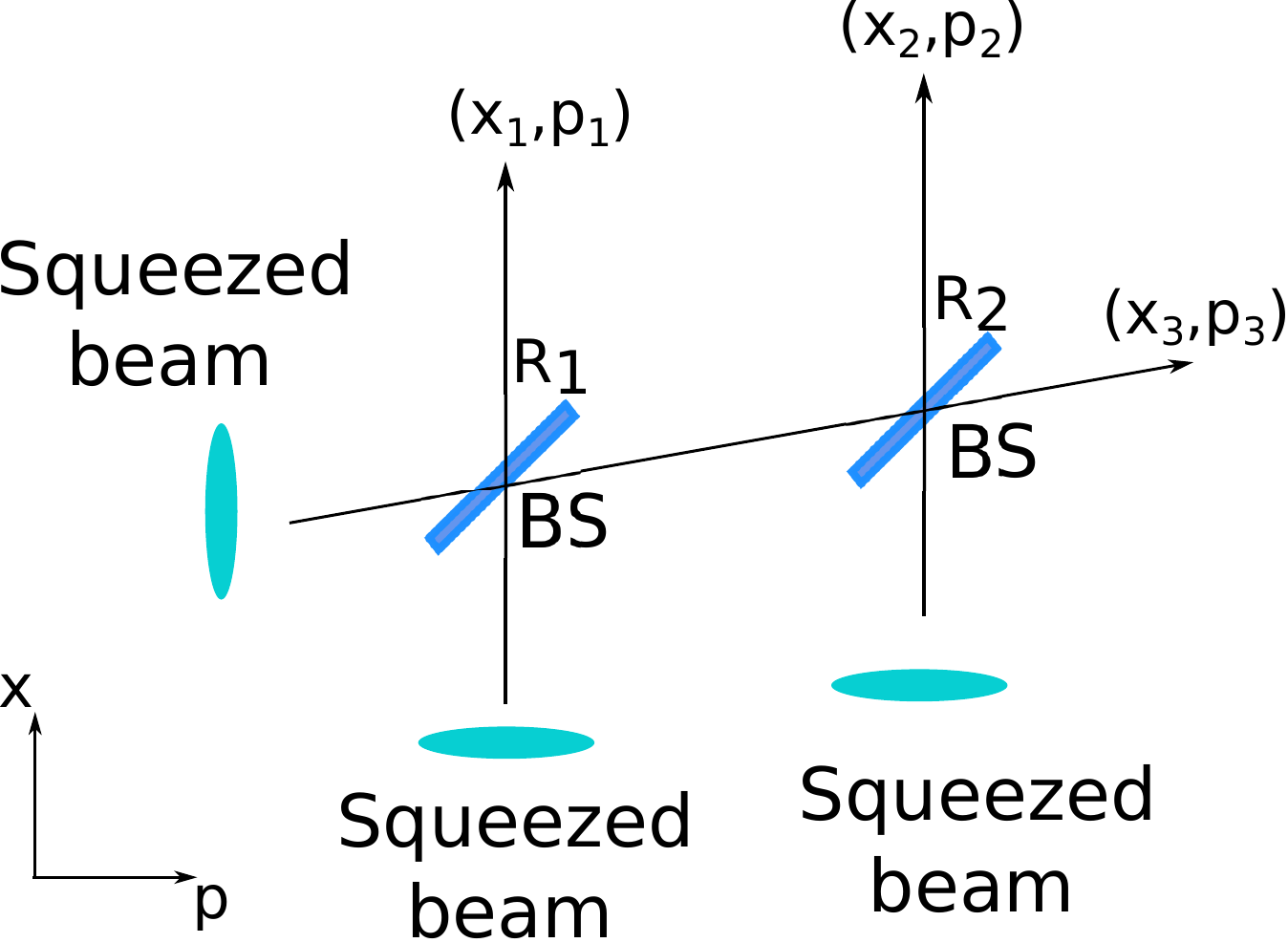}

\caption{Generation of the tripartite-entangled CV GHZ state. The configuration
uses three squeezed-vacuum inputs and two beam splitters (BS) with
reflectivities $R_{1}=1/3$ and $R_{2}=1/2$. The $x_{i}$ and $p_{i}$
are the two orthogonal quadrature-phase amplitudes of the spatially
separated optical modes $i$ $(i=1,2,3)$. \label{fig:tripartite_ent_GHZ}}
\end{figure}

\begin{figure}
\includegraphics[width=0.7\columnwidth]{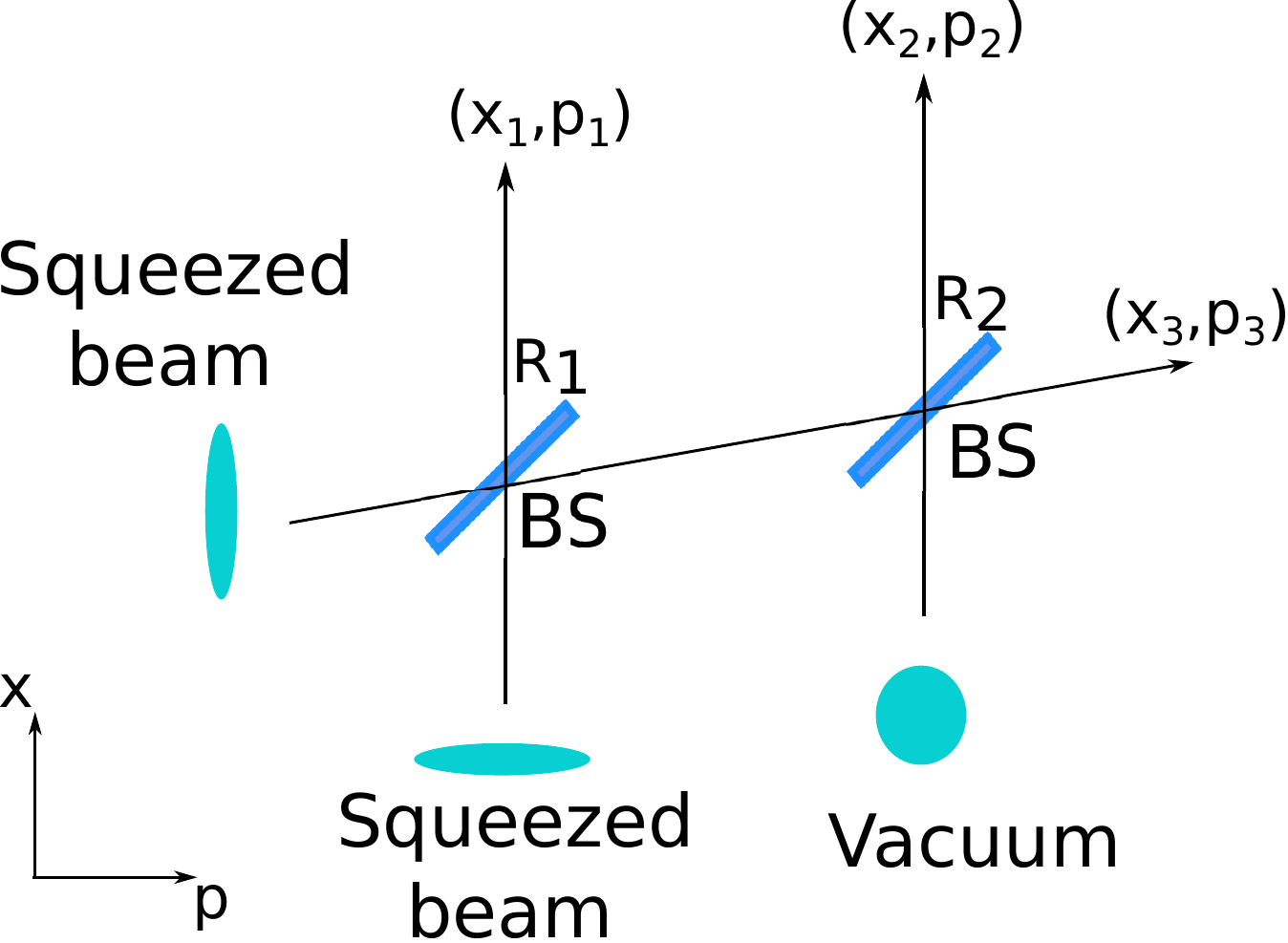}

\caption{Generation of the tripartite entangled EPR-type state. The configuration
uses two squeezed-vacuum inputs, a coherent-vacuum input, and two
beam splitters BS with reflectivities $R_{1}=R_{2}=1/2$. The $x_{i}$
and $p_{i}$ are the two orthogonal quadrature-phase amplitudes of
the spatially separated optical modes $i$ $(i=1,2,3)$.\label{fig:tripartite_ent_cvEPR}}
\end{figure}

\textcolor{black}{This scheme can be extended to generate genuine
tripartite polarization entanglement. Genuine CV tripartite entanglement
in the quadratures is first created in an optical setup involving
squeezed vacuums and beam splitters, as shown in Figs. \ref{fig:tripartite_ent_GHZ}
and \ref{fig:tripartite_ent_cvEPR}. The three entangled modes from
the outputs of these beam splitters are horizontally polarized. Each
of these modes is subsequently mixed with a bright coherent beam with
vertical polarization using a polarizing beam splitter. At each site
$i=1,2,3$ prior to mixing, one can define pairs of orthogonally polarized
modes (with annihilation operators $a_{Hi}$, $a_{V,i}$). The choice
of polarizer angle determines which Stokes observable is measured,
after a number difference is taken. The final readout is given as
a difference current. After the mixing, the genuine CV entanglement
has been transformed into genuine tripartite polarization entanglement,
as illustrated in Figure 3.}

\textcolor{black}{To verify the tripartite polarization entanglement,
we consider the sum inequality 
\begin{align}
 & \Delta^{2}\left[S_{y,1}+h\left(S_{y,2}+S_{y,3}\right)\right]+\Delta^{2}\left[S_{z,1}+g\left(S_{z,2}+S_{z,3}\right)\right]\nonumber \\
 & \geq2\text{min}\{\alpha_{v}^{2}+2\left|gh\right|\alpha_{v}^{2},\left|gh\right|\alpha_{v}^{2}+\alpha_{v}^{2}\left|1+gh\right|\}\,,\label{eq:stokes_sum}
\end{align}
where $\alpha_{v}$ is the coherent amplitude of the vertically polarized
coherent beam.} Here, we assume the limit of large coherent amplitude
$\alpha_{v}$ and Poissonian fluctuations, which allows the linearization
of the mode amplitude $a_{V}$.

\textcolor{black}{The variances are 
\begin{align}
\Delta^{2}\left[S_{y,1}\!+\!h\left(S_{y,2}\!+\!S_{y,3}\right)\right] & =\alpha_{v}^{2}\Delta^{2}\left[P_{H,1}\!+\!h\!\left(P_{H,2}\!+\!P_{H,3}\right)\right]\nonumber \\
\Delta^{2}\left[S_{z,1}\!+\!g\left(S_{z,2}\!+\!S_{z,3}\right)\right] & =\alpha_{v}^{2}\Delta^{2}[X_{H,1}\!+\!g\!\left(X_{H,2}\!+\!X_{H,3}\right)]\,.\label{eq:stokes_var}
\end{align}
Here, $S_{x,k}$, $S_{y,k}$ and $S_{z,k}$ are the Stokes operators
defined in (\ref{eq:stokes_op}) for each mode pair at site $k$.
$X_{H,k}$ and $P_{H,k}$ are the $X$ and $P$ quadratures for beam
$k$: $P_{H,k}=a_{H,k}^{\dagger}e^{i\theta}+a_{H,k}e^{-i\theta}$
and $X_{H,k}=ia_{H,k}e^{-i\theta}-ia_{H,k}^{\dagger}e^{i\theta}$.}
\textcolor{black}{The $h$ and $g$ are gain factors defined in the
criteria, where we take $h_{1}=1$, $h_{2}=h_{3}=h$ and $g_{1}=1$,
$g_{2}=g_{3}=g$. Note that the commutation relations satisfied by
these Stokes operators are $\left[S_{x},S_{y}\right]=2iS_{z}$, which
differ from the spin commutation relations by a factor of $2$. As
a result, the sum and product inequalities below have an extra factor
of $2$ compared to the sum and product inequalities in Eqs. (\ref{eq:criterion-1})
and (\ref{eq:criterion-2}) respectively. With these variances, the
sum inequality Eq. (\ref{eq:criterion-1}) and the product inequality
Eq. (\ref{eq:criterion-2}) are respectively transformed into a continuous-variable
genuine tripartite entanglement sum and product criterion, according
to 
\begin{align}
 & \frac{\Delta^{2}\left[S_{y,1}+h\left(S_{y,2}+S_{y,3}\right)\right]+\Delta^{2}\left[S_{z,1}+g\left(S_{z,2}+S_{z,3}\right)\right]}{2\alpha_{v}^{2}\text{min}\{1+2\left|gh\right|,\left|gh\right|+\left|1+gh\right|\}}\nonumber \\
\nonumber \\
 & =\frac{\Delta^{2}\!\left[X_{H,1}\!+\!g\left(\!X_{H,2}\!+\!X_{H,3}\right)\right]\!+\!\Delta^{2}\!\left[P_{H,1}\!+\!h\left(\!P_{H,2}\!+\!P_{H,3}\right)\right]}{2\text{min}\{1+2\left|gh\right|,\left|gh\right|+\left|1+gh\right|\}}\nonumber \\
 & \geq1\,\label{eq:stokes_ent_sum}
\end{align}
}and

\textcolor{black}{{} 
\begin{align}
 & \frac{\Delta\left[S_{y,1}+h\left(S_{y,2}+S_{y,3}\right)\right]\Delta\left[S_{z,1}+g\left(S_{z,2}+S_{z,3}\right)\right]}{\text{min}\{\alpha_{v}^{2}+2\left|gh\right|\alpha_{v}^{2},\left|gh\right|\alpha_{v}^{2}+\alpha_{v}^{2}\left|1+gh\right|\}}\nonumber \\
\nonumber \\
 & =\frac{\Delta\left[X_{H,1}\!+\!g\left(X_{H,2}\!+\!X_{H,3}\right)\right]\Delta\left[P_{H,1}\!+\!h\left(P_{H,2}\!+\!P_{H,3}\right)\right]}{\text{min}\{1+2\left|gh\right|,\left|gh\right|+\left|1+gh\right|\}}\nonumber \\
 & \geq1\,.\label{eq:stokes_ent_prod}
\end{align}
Hence, any CV genuine tripartite quadrature entanglement then implies
genuine tripartite polarization entanglement.}

\begin{figure}
\includegraphics[width=1\columnwidth]{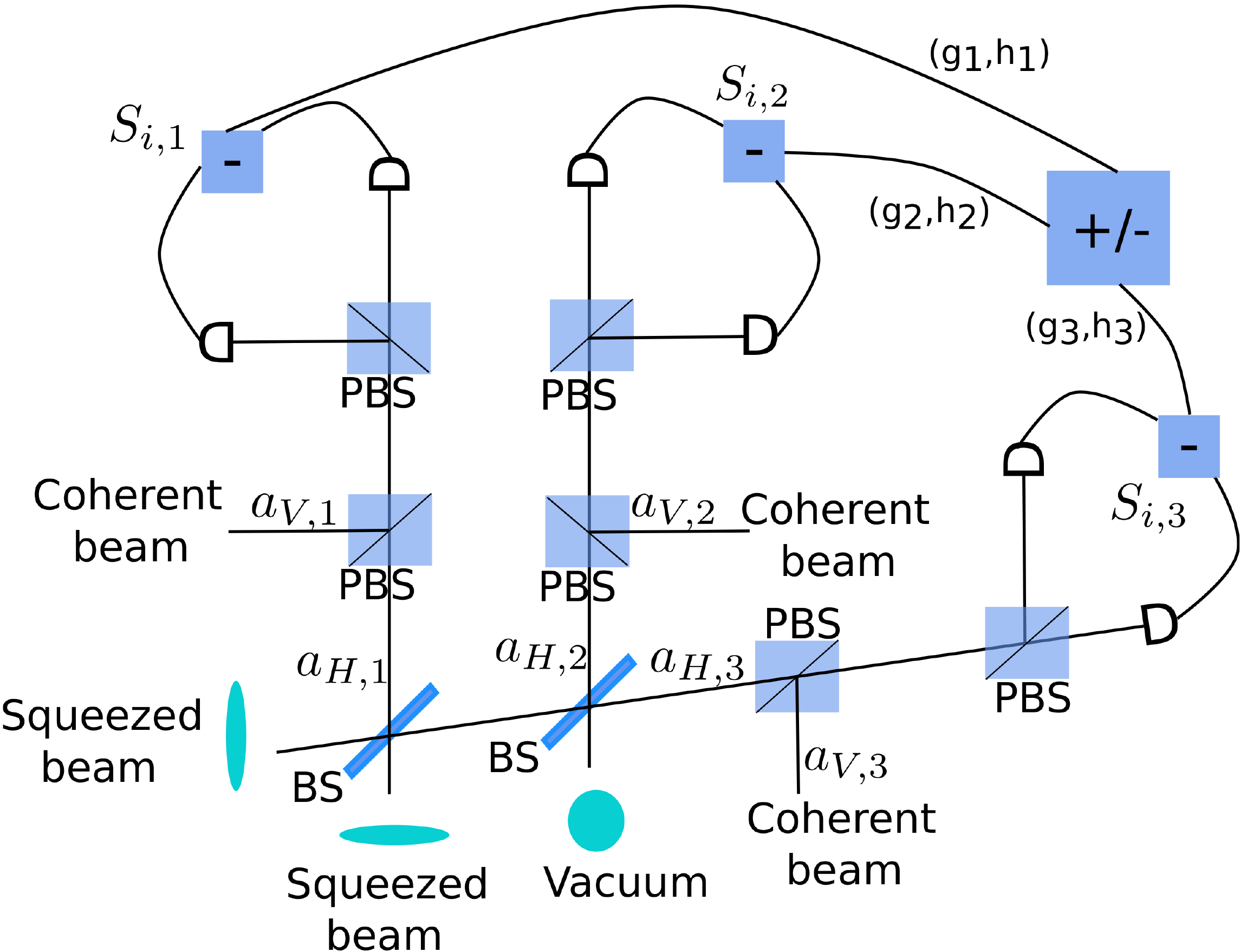}

\caption{\textcolor{black}{The experimental setup to generate genuine tripartite
polarization entanglement from genuine tripartite CV entanglement.
In this schematic diagram, an EPR-type genuine tripartite-entanglement
is generated as shown in Figure \ref{fig:tripartite_ent_cvEPR}. The
outputs are mixed with coherent fields, as described in the text.
The $S_{i,k}$ denotes the polarisation $S_{x,k}$, $S_{y,k}$ or
$S_{z,k}$ for the site $k$ ($k=1,2,3$). The $\left(g_{k},h_{k}\right)$
are the gains used in the criteria, and are introduced in the final
currents. By using a third squeezed input state at the second beam
splitter instead of the vacuum input, the CV GHZ genuine tripartite
entanglement (refer Figure \ref{fig:tripartite_ent_GHZ}) }can be
transformed into an equivalent genuine tripartite polarization entanglement.
Alternatively, by using only one squeezed input, one may transfer
the genuine tripartite entanglement depicted in Figure \ref{fig:tripartite_ent_BS}.\label{fig:bowen}}
\end{figure}

\textcolor{black}{There are two types of states that show genuine
tripartite entanglement in the quadratures. These are the CV GHZ and
CV EPR-type states, defined in Refs. \cite{vanLoock_PRA2003} and
\cite{PhysRevA.90.062337}, and illustrated in Figs. \ref{fig:tripartite_ent_GHZ}
and \ref{fig:tripartite_ent_cvEPR} respectively. It has been shown
previously that these two states violate both the quadrature sum inequality
in Eq. (\ref{eq:stokes_ent_sum}) and the product inequality in Eq.
(\ref{eq:stokes_ent_prod}) with specific values for the gains, $g_{1}=h_{1}=1$
and $g_{i>1}=g,\,h_{i>1}=h$ \cite{PhysRevA.90.062337}. The gains
$g,h$ are chosen such that the variance sum and product are minima,
and are given in Table \ref{tab:gains}. With these gain values, as
shown in Ref. \cite{PhysRevA.90.062337}, Criteria 1 and 3 are always
violated for any nonzero squeezing of the squeezed vacuum inputs,
implying the presence of genuine tripartite entanglement. The inequalities
of Criteria 1 and 3 are also useful in showing genuine tripartite
entanglement. The optimal gains for these inequalities can be found
in Ref. \cite{PhysRevA.90.062337}.}

\begin{figure}
\includegraphics[width=0.7\columnwidth]{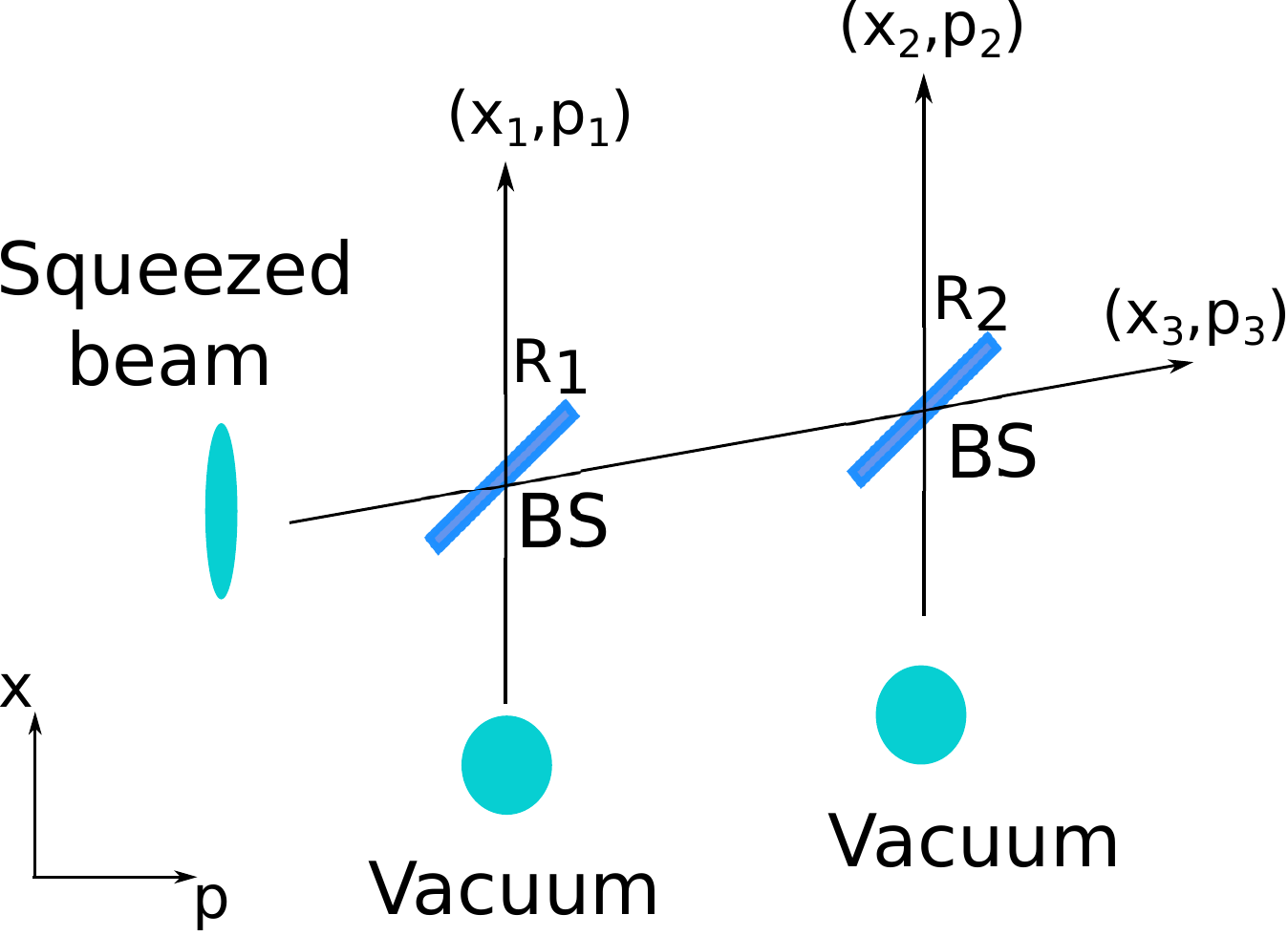}

\caption{Generation of tripartite entanglement using a squeezed vacuum beam
with squeezing \textcolor{black}{along the $P$ (or $X$) quadrature.}\textcolor{red}{{}
}All other beam splitter ports have vacuum inputs. The reflectivity
for the first beam splitter is $R_{1}=1/3$ and $R_{2}=1/2$ for the
second beam splitter. \textcolor{black}{A calculation of the genuine
tripartite entanglement generated from this configuration is given
in Appendix 3. }\label{fig:tripartite_ent_BS}}
\end{figure}

\textcolor{black}{Genuine tripartite entanglement is also created
using a third configuration involving only one squeezed input, shown
in Fig. \ref{fig:tripartite_ent_BS}. Normally, two squeezed vacuum
inputs are combined across a beam splitter to create strong EPR-correlations
between the output modes \cite{Reid_colloqium_RMP2009,Aoki_PRL2003}.
It is also possible to create EPR-entangled modes, using only one
squeezed vacuum input \cite{Reid_PRA1989}. While the EPR correlations
are weaker, the entanglement is sufficiently strong that a subsequent
beam-splitter interaction with a non-squeezed vacuum input can create
genuine tripartite entanglement. A summary of this calculation is
given in the Appendix 3, where we show how the entanglement that is
generated can be detected by Criterion 5 of Ref. \cite{PhysRevA.90.062337}
with the gains $h=-1/2$ and $g=1/2$. This tripartite entanglement
is not sufficiently strong to generate tripartite EPR-steering correlations,
but can be transformed into genuine tripartite polarization-entanglement
using the configuration of Fig. \ref{fig:bowen}. The spin sum-inequality
given by Criterion 3 is then useful to detect the genuine tripartite
entanglement. }

\begin{table}
\begin{tabular}{|c|c|c|c|c|}
\hline 
\multirow{2}{*}{\textbf{r}} & \multicolumn{2}{c|}{\textbf{CV GHZ}} & \multicolumn{2}{c|}{\textbf{CV EPR}}\tabularnewline
 & \multicolumn{1}{c}{$h$} & $g$ & \multicolumn{1}{c}{$h$} & $g$\tabularnewline
\hline 
\hline 
0 & 0 & 0 & 0 & 0\tabularnewline
\hline 
0.25 & 0.36 & -0.27 & 0.33 & -0.33\tabularnewline
\hline 
0.50 & 0.68 & -0.40 & 0.54 & -0.54\tabularnewline
\hline 
0.75 & 0.86 & -0.46 & 0.64 & -0.64\tabularnewline
\hline 
1.00 & 0.95 & -0.49 & 0.68 & -0.68\tabularnewline
\hline 
1.50 & 0.99 & -0.50 & 0.70 & -0.70\tabularnewline
\hline 
2.00 & 1.00 & -0.50 & 0.70 & -0.70\tabularnewline
\hline 
\end{tabular}

\caption{Values of the gains $g$ and $h$ that minimize the variance sum and
product in Criteria 1 and 3. \label{tab:gains}}
\end{table}

\subsubsection{Validity of van Loock-Furusawa type inequalities}

Criteria 2, 4 and 10 can be used for any value of $\langle J_{z,k}\rangle$
and $\Delta^{2}J_{z,k}$ i.e. without the assumption $\left|\langle J_{z,k}\rangle\right|\rightarrow\infty$
and $\Delta^{2}J_{z,k}/\left|\langle J_{z,k}\rangle\right|\le1$.
Reference \cite{PhysRevA.90.062337} derives continuous-variable van
Loock-Furusawa Criteria that are useful to demonstrate the genuine
tripartite entanglement of the CV GHZ and EPR states discussed in
Sections V.A and depicted in Figs. \ref{fig:tripartite_ent_GHZ} and
\ref{fig:tripartite_ent_cvEPR}. These are given as Criteria 1 and
2 (Eqs. 14 and 15 respectively) in Ref. \cite{PhysRevA.90.062337}.
In Section V.A, a correspondence between the spin and continuous variable
cases is given. Based on that correspondence, we see that Criteria
2 and 4 are useful to demonstrate the genuine tripartite entanglement
of the spin systems depicted in Fig. \ref{fig:bowen}, without extra
assumptions.

\subsection{\textcolor{black}{\normalsize{}{}Tripartite entanglement of atomic
ensembles \label{subsec:atomic_ensembles}}}

\textcolor{black}{Tripartite entanglement can also be created among
three atomic ensembles by successively passing polarized light through
the ensembles. Here, we outline a generalization of the scheme of
Julsgaard et al. that creates bipartite entanglement between two atomic
ensembles \cite{Julsgaard:2001aa}. The observables for the atomic
ensembles are the collective Schwinger spins defined by the operators:
\begin{align}
\hat{J}_{x} & =\frac{1}{2}\left(a_{+}^{\dagger}a_{+}-a_{-}^{\dagger}a_{-}\right)\nonumber \\
\hat{J}_{y} & =\frac{1}{2}\left(a_{+}^{\dagger}a_{-}e^{i\theta}+a_{-}^{\dagger}a_{+}e^{-i\theta}\right)\nonumber \\
\hat{J}_{z} & =\frac{1}{2}\left(ia_{-}^{\dagger}a_{+}e^{-i\theta}-ia_{+}^{\dagger}a_{-}e^{i\theta}\right)\,,\label{eq:spin_operators}
\end{align}
which satisfy the commutation relation $\left[\hat{J}_{x},\hat{J}_{y}\right]=i\hat{J}_{z}$.
Here, $a_{+},a_{-}$ are the operators for spin-up and spin-down along
the spin-$x$ axis, respectively. We label the operators for each
ensemble by the subscript $k$ ($k=1,2,3$).}

\textcolor{black}{Firstly, three atomic ensembles are prepared such
that the mean collective spins for these atomic ensembles are pointing
along the $x$-axis: $J_{x1}=-2J_{x2}=-2J_{x3}=J_{x}$. A linearly
polarized light along the $x$-axis is then applied to the ensembles.
The light-spin interaction is given by the Hamiltonian $H_{\text{int}}=\omega\hat{S}_{z}\hat{J}_{z123}$,
where $\hat{J}_{z123}=\hat{J}_{z1}+\hat{J}_{z2}+\hat{J}_{z3}$. The
light variable then evolves in terms of the inputs to give an output
of 
\begin{align}
\hat{S}_{y}^{\text{out}} & =\hat{S}_{y}^{\text{in}}+\alpha\hat{J}_{z123}\,,\label{eq:sy_out}
\end{align}
while the spin variables evolve as 
\begin{align}
\hat{J}_{y1}^{\text{out}} & =\hat{J}_{y1}^{\text{in}}+\beta\hat{S}_{z}\nonumber \\
\hat{J}_{y2}^{\text{out}} & =\hat{J}_{y2}^{\text{in}}-\frac{1}{2}\beta\hat{S}_{z}\nonumber \\
\hat{J}_{y2}^{\text{out}} & =\hat{J}_{y2}^{\text{in}}-\frac{1}{2}\beta\hat{S}_{z}\,.\label{eq:jy_out}
\end{align}
By measuring $\hat{S}_{y}^{\text{out}}$, $\hat{J}_{z1}+\hat{J}_{z2}+\hat{J}_{z3}$
can be inferred. Also, $\hat{J}_{y1}+\hat{J}_{y2}+\hat{J}_{y3}$ can
be measured using another light pulse without affecting the measured
value of $\hat{J}_{z1}+\hat{J}_{z2}+\hat{J}_{z3}$. This is possible
because $\left[\hat{J}_{z1}+\hat{J}_{z2}+\hat{J}_{z3},\hat{J}_{y1}+\hat{J}_{y2}+\hat{J}_{y3}\right]=0$.
Hence, the quantity $\Delta^{2}\left(\hat{J}_{z1}+\hat{J}_{z2}+\hat{J}_{z3}\right)+\Delta^{2}\left(\hat{J}_{y1}+\hat{J}_{y2}+\hat{J}_{y3}\right)$
can be arbitrarily small. Using the sum inequality Eq. (\ref{eq:criterion-1})
and product inequality Eq. (\ref{eq:criterion-2}) with gain values
$g_{i}=h_{i}=1$, $\left(i=1,2,3\right)$, a genuine tripartite entanglement
is certified among the atomic ensembles if $\Delta^{2}\left(\hat{J}_{z1}+\hat{J}_{z2}+\hat{J}_{z3}\right)+\Delta^{2}\left(\hat{J}_{y1}+\hat{J}_{y2}+\hat{J}_{y3}\right)<2J_{x}$
for the sum inequality and $\Delta\left(\hat{J}_{z1}+\hat{J}_{z2}+\hat{J}_{z3}\right)\Delta\left(\hat{J}_{y1}+\hat{J}_{y2}+\hat{J}_{y3}\right)<J_{x}$
for the product inequality.}

\subsection{Entangled Bose-Einstein condensate clouds \label{subsec:bec_cloud}}

\textcolor{black}{In the experiment of Kunkel et al. \cite{Kunkel413},
a $^{87}\text{Rb}$ Bose-Einstein condensate is first generated in
the magnetic substate $m_{F}=0$ of the $F=1$ hyperfine manifold,
before a spin-squeezing operation coherently populates the $m_{F}=\pm1$
states and entangles all the atoms in the condensate. The condensate
is then allowed to expand into three spatially separated partitions.
The tripartite entanglement among these partitions is verified by
measuring $\hat{F}_{0,k}$ and $\hat{F}_{\pi/2,k}$ for each partition
$k$, where $\hat{F}_{\phi,k}=\left[\left(\hat{a}_{+1}^{\dagger}+\hat{a}_{-1}^{\dagger}\right)\hat{a}_{0}e^{i\phi}+h.c.\right]/\sqrt{2}$,
$\hat{a}_{j}^{\dagger}$ is the creation operator for a state $m_{F}=j$.
These operators satisfy the commutation relation $\left[\hat{F}_{0,k},\hat{F}_{\pi/2,k}\right]=2i\hat{N}_{k}$,
where $\hat{N}_{k}$ is the number operator for the partition $k$.
By applying $\pi/2$ pulses and rotations, these observables are measured
by reading out the population difference between the states $m_{F}=\pm1$.
If the number of atoms in group $m_{F}=0$ is large, then the measurement
becomes similar to a homodyne detection of the amplitudes ($\left(\hat{a}_{+1}^{\dagger}+\hat{a}_{-1}^{\dagger}\right)e^{i\phi}+h.c.$)
associated with the atoms of each of the partitions, carried out with
the second larger group of atoms (given by $\hat{a}_{0}$) acting
as the local oscillator, as explained in Refs. \cite{Ferris_PRA2008,Gross_Oberthaler_Nature2011}.
More generally, spin relations must be used. In the atomic experiment
of Kunkel et al., the genuine $N$-partite entanglement (up to $N=5$)
mutually shared among the clouds is certified using criteria similar
to that derived in Ref. \cite{PhysRevA.90.062337}, for quadrature
phase amplitudes, but properly accounting for the spin and number
operators that apply in this case. }

\textcolor{black}{In another experiment based on the two hyperfine
states $|1\rangle=|F=1,m_{F}=-1\rangle$ and $|2\rangle=|F=2,m_{F}=1\rangle$
{} of a $^{87}\text{Rb}$ BEC, Fadel et al. \cite{Fadel409} prepare
the system in an atomic spin-squeezed state, and allow the condensate
to expand into two separated partitions (which we denote $A$ and
$B$). This creates a bipartite entanglement between the two clouds,
which is detected using the entanglement criterion \cite{Giovannetti_PRA2003,Fadel409}
\begin{align}
 & \Delta\left(g_{z}S_{z,A}+S_{z,B}\right)\Delta\left(g_{y}S_{y,A}+S_{y,B}\right)\nonumber \\
 & <\frac{1}{2}\left(\left|g_{z}g_{y}\right|\left|\langle S_{x,A}\rangle\right|+\left|\langle S_{x,B}\rangle\right|\right)\,.\label{eq:fadel_ent}
\end{align}
Here, $S_{z,A/B}$ and $S_{y,A/B}$ are the collective Schwinger spin
operators \cite{Hu_PRA2015,Pezze_RMP2018} along the $z$- axis and
$y$-axis respectively, for partition $A/B$. Explicitly, the collective
spin operator $S_{z,A/B}$ is given as the number difference 
\begin{align}
S_{z,A/B} & =\frac{1}{2}\left(N_{z,A/B}^{1}-N_{z,A/B}^{2}\right)\,,\label{eq:S_z,A/B}
\end{align}
where $N_{z,A/B}^{1}$ and $N_{z,A/B}^{2}$ are the number of atoms
in the internal spin states $|1\rangle$ and $|2\rangle$ respectively,
along the spin $z$-axis, for partition $A/B$. The collective spin
operators along the $y$-axis $S_{y,A/B}$ are defined accordingly
following Eq. (\ref{eq:spin_operators}), but noting the switching
of the labels $x,y,z$. Other proposals exist to create a similar
bipartite entanglement that can be detected using a similar spin criterion
\cite{He_EPR_strategies_PRL2011,He_EPRsteering_double_wellBEC_PRA2012,Opanchuk_double_wellBEC_PRA2012}.}

\textcolor{black}{The experiment of Fadel et al. observed bipartite
entanglement and EPR steering, but did not investigate tripartite
entanglement. It is likely however that one could detect a genuine
tripartite entanglement for clouds generated by further splitting
the BEC. This would seem possible, given the result obtained in the
Appendix 3 and depicted in Fig. \ref{fig:tripartite_ent_BS}, where
tripartite entanglement is generated using only one squeezed input,
followed by a sequence of splitting of the modes using beam splitter
interactions. This works, because entangled modes can be created from
a beam splitter with only one squeezed vacuum input \cite{Reid_PRA1989}}.
The tripartite entanglement created in the three modes of Fig. \textcolor{black}{\ref{fig:tripartite_ent_BS}}
can be detected \textcolor{black}{using the Criterion 5 of Ref. \cite{PhysRevA.90.062337}
with the gains $h=-1/2$ and $g=1/2$. If one considers transforming
into an equivalent tripartite entanglement in the Schwinger operators,
then the suitable criterion would be Criterion 3 in Eq. (\ref{eq:criterion-2})
with the gains $h=-1/2$ and $g=1/2$.}

A realization of a beam splitter interaction for the BEC can be obtained
in several ways. An analogy of optical beam splitters with the splitting
of a condensate (which is envisaged to be a realization of the final
beam splitter of Fig. \ref{fig:tripartite_ent_BS}) is explained in
Ref \cite{PhysRevLett.116.080402}.\textcolor{black}{{} The splitting
into two modes is described by the interaction Hamiltonian 
\begin{align}
H_{I+} & =e^{i\phi}a_{+}^{\dagger}a_{+0}+e^{-i\phi}a_{+}a_{+0}^{\dagger}\,,\label{eq:BS_hamiltonian}
\end{align}
where $a_{+},a_{+0}$ are the annihilation operators for modes labelled
$A_{+}$ and $A_{+,0}$ respectively, and $\phi$ is the phase difference
between these two modes. The transformation is equivalent to the beam
splitter relations 
\begin{align}
a_{+,\text{out}} & =a_{+}\cos\tau-ie^{i\phi}a_{+0}\sin\tau\nonumber \\
a_{+0,\text{out}} & =a_{+0}\cos\tau-ie^{-i\phi}a_{+}\sin\tau\,,\label{eq:transform}
\end{align}
where $\tau$ is the interaction time and $a_{+,\text{out}}=a_{+}\left(\tau\right)\,,a_{+0,\text{out}}=a_{+0}\left(\tau\right)$.
One can adjust the effective transmission to reflection ratio by adjusting
the interaction time between the two modes.}

\textcolor{black}{We thus consider two separated clouds $A$ and $B$
that show spin entanglement with respect to the difference operators
$g_{z}S_{z,A}+S_{z,B}$ and $g_{y}S_{y,A}+S_{y,B}$ so that the criterion
of Eq. (\ref{eq:fadel_ent}) is satisfied. These two clouds are analogous
to the entangled outputs after the first beam splitter $BS$ of the
configuration shown in Fig. \ref{fig:tripartite_ent_BS}. Each cloud
is identified with Schwinger spin observables. For example, $S_{z,A}$
and $S_{y,A}$ are measurements that can be made on cloud $A$, where
$S_{z,A}=\frac{1}{2}\left(a_{+}^{\dagger}a_{+}-a_{-}^{\dagger}a_{-}\right)$
and $a_{+}$, $a_{-}$ correspond to the two atomic levels. To generate
the tripartite entanglement, the system $A$ is transformed according
to a beam splitter interaction (splitting) modeled as Eq. (\ref{eq:BS_hamiltonian}).
Since the splitting is insensitive to the internal spin degrees of
freedom, there is a similar independent interaction for $a_{-}$.
The outputs of $a_{\pm1}$ and $a_{\pm2}$ are then spatially separated,
so that three separate clouds are created, labelled $A1$, $A2$ and
$B$, these being analogous to the three outputs of the configuration
of Fig. \ref{fig:tripartite_ent_BS}. The final Schwinger operators
at $A_{1}$ and $A_{2}$ are defined by the $a_{\pm1}$ at $A_{1}$,
and the $a_{\pm2}$ at $A_{2}$. The different Schwinger components
can be measured using Rabi rotations or equivalent \cite{Gross_Oberthaler_Nature2011,Fadel409}.
The calculation carried out in Appendix 5 predicts a tripartite entanglement
between the three clouds that could be detected by Criteria 1 and
3. Using Eqs. (\ref{eq:S_zA1}) and (\ref{eq:S_zA2}) in Appendix
5, the inequality of Criterion 3 is then 
\begin{align}
 & \Delta\!\left[g_{z}\!\left(S_{z,A1}\!+\!S_{z,A2}\right)\!+\!S_{z,B}\right]\!\Delta\!\left[g_{y}\!\left(S_{y,A1}\!+\!S_{y.A2}\right)\!+\!S_{y,B}\right]\nonumber \\
 & \geq\frac{1}{2}\text{min}\{\left|g_{z}g_{y}\right|\left|\langle S_{x,A1}\rangle+\langle S_{x,A2}\rangle\right|+\left|\langle S_{x,B}\rangle\right|\,,\nonumber \\
 & \;\;\left|(g_{z}g_{y}\langle S_{x,A1}\rangle+\langle S_{x,B}\rangle)\right|+\left|g_{z}g_{y}\right|\left|\langle S_{x,A2}\rangle\right|\}\,.\label{eq:criterion3_BECsplitting}
\end{align}
The violation of this inequality implies genuine tripartite entanglement.
We show in Appendix 5 that, assuming the number of atoms is large,
$S_{z,A1}+S_{z,A2}\approx S_{z,A}$, $S_{y,A1}+S_{y,A2}\approx S_{y,A}$,
and $S_{x,A1}+S_{x,A2}\approx S_{x,A}$. The criterion for genuine
tripartite entanglement will therefore be satisfied if there is sufficient
entanglement as measured by the bipartite criterion given in Eq. (\ref{eq:fadel_ent}).
Assuming $S_{x,A}$ and $S_{x,B}$ correspond to the Bloch vectors,}\textcolor{red}{{}
}\textcolor{black}{with the directions of axes being chosen to ensure
$\langle S_{x,A}\rangle$ and $\langle S_{x,B}\rangle$ are positive,
we see that{} the beam splitter transformation (refer Appendix 5)
ensures the signs of $S_{x,A1}$ and $S_{x,A2}$ are also positive.
The right-side of the inequality is then either precisely that given
by the right-side of Eq. (\ref{eq:fadel_ent}) (if $g_{z}g_{y}>0$),
or is less than this value (if $g_{z}g_{y}<0$). }

\begin{figure}
\includegraphics[width=0.99\columnwidth]{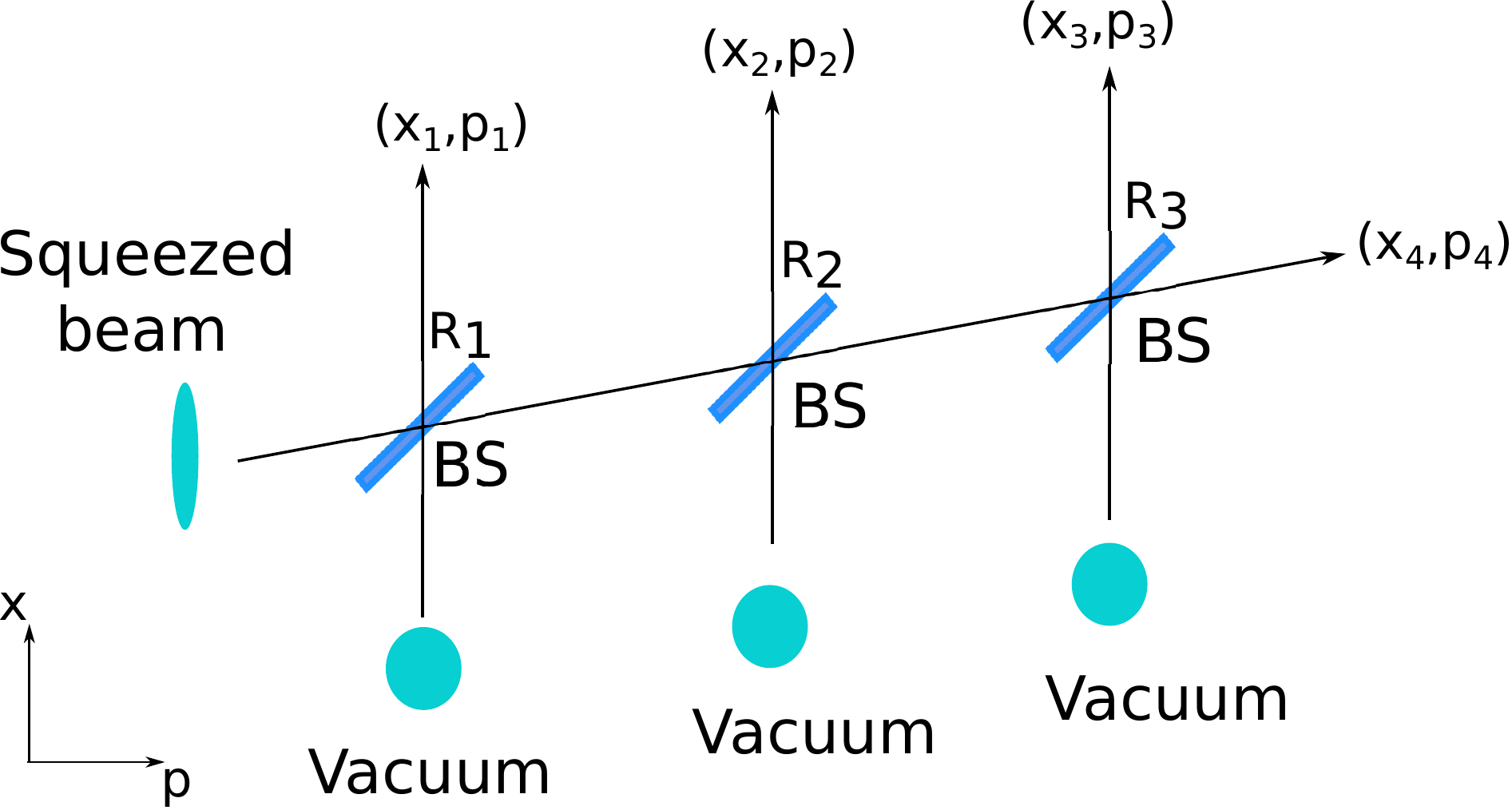}

\caption{Generation of four-partite entanglement using a squeezed beam \textcolor{black}{along
the $P$ (or $X$) quadrature.}\textcolor{red}{{} }All other beam
splitter ports have vacuum inputs. The reflectivity for the first
beam splitter $R_{1}=1/4$, $R_{2}=1/3$ for the second beam splitter
and $R_{3}=1/2$ for the third beam splitter. \textcolor{black}{{} }\label{fig:fourpartite_ent_BS}}
\end{figure}

We note from the results reported in \textcolor{black}{Refs. \cite{vanLoock_PRA2003,Armstrong_Nature2012,PhysRevA.90.062337}}\textcolor{blue}{{}
}that we can generate $N$-partite entangled states ($N>3)$ by successive
use of beam splitters with vacuum inputs, once an initial entangled
state is created from two squeezed inputs or some other means. This
has been implemented for a BEC by Ref. \cite{Kunkel413} (for $N=5$).
We show in the Appendix 4 that we can also create genuinely $4-$partite
entangled states from a single squeezed input (refer Fig. \ref{fig:fourpartite_ent_BS}),
followed by multiple beam splitter combinations and vacuum inputs
(with no squeezing). This may provide an avenue (using successive
splittings) for the generation of multi-partite entanglement in experiments
such as Ref. \cite{Fadel409}.

\subsection{{\normalsize{}{}Remarks on the applicability and validity of genuine
multipartite entanglement criteria}}

In this subsection, we discuss the applicability and validity of the
genuine multipartite entanglement criteria in the limit of large mean
spins and small fluctuations where $\Delta^{2}J_{z,k}/\left|\langle J_{z,k}\rangle\right|\leq1$
and hence $\Delta J_{z,k}/|\langle J_{z,k}\rangle|\rightarrow0$ as
$|\langle J_{z,k}\rangle|\rightarrow\infty$.

This limiting case justifies the application of Criterion 1 in Section
V.A, because in that system the fields denoted $a_{V,i}$ are assumed
to be intense coherent fields (see below). It also justifies application
of the continuous variable Criteria derived in Ref. \cite{PhysRevA.90.062337}
to experiments where the quadrature phase amplitude of each mode is
measured via homodyne detection. This includes the approach summarized
at the beginning of Section V.C. The local oscillators used in the
homodyne detection correspond to second field modes $a_{V,i}$, assumed
to be intense classical fields with fixed amplitude. To prove these
cases directly, we note that the measurement of the number difference
at the beam splitter of each homodyne detection corresponds to that
of a Schwinger operator: We define 
\begin{align}
J_{z,k} & =(a_{H,k}^{\dagger}a_{H,k}-a_{V,k}^{\dagger}a_{V,k})/2\nonumber \\
J_{x,k} & =(a_{H,k}^{\dagger}a_{V,k}+a_{V,k}^{\dagger}a_{H,k})/2\nonumber \\
J_{y,k} & =(ia_{V,k}^{\dagger}a_{H,k}-ia_{H,k}^{\dagger}a_{V,k})/2\,,\label{eq:stokes_op-1-1}
\end{align}
Taking the limit where field $a_{V,k}$ is the coherent intense local
oscillator field, approximated by a classical amplitude $\alpha e^{-i\theta}$
and so that $\langle a_{H,k}^{\dagger}a_{H,k}\rangle\ll\langle a_{V,k}^{\dagger}a_{V,k}\rangle$,
we define 
\begin{align}
X_{k} & =\sqrt{2}J_{x,k}/\langle J_{z,k}\rangle^{1/2}\nonumber \\
 & =(a_{H,k}^{\dagger}a_{V,k}+a_{V,k}^{\dagger}a_{H,k})/\langle(a_{H,k}^{\dagger}a_{H,k}-a_{V,k}^{\dagger}a_{V,k}\rangle^{1/2}\nonumber \\
\end{align}
\begin{align}
P_{k} & =-\sqrt{2}J_{y,k}/\langle J_{z,k}\rangle^{1/2}\nonumber \\
 & =-(ia_{V,k}^{\dagger}a_{H,k}-ia_{H,k}^{\dagger}a_{V,k})/\langle(a_{H,k}^{\dagger}a_{H,k}-a_{V,k}^{\dagger}a_{V,k}\rangle^{1/2}
\end{align}
Replacing $a_{V,k}$ with a classical amplitude gives 
\begin{align}
X_{k} & =(\alpha e^{i\theta}a_{H,k}+a_{H,k}^{\dagger}\alpha e^{-i\theta})/\langle2J_{z,k}\rangle^{1/2}\nonumber \\
 & =e^{i\theta}a_{H,k}+a_{H,k}^{\dagger}e^{-i\theta}\nonumber \\
P_{k} & =(ia_{H,k}^{\dagger}\alpha e^{-i\theta}-i\alpha e^{i\theta}a_{H,k})/\langle2J_{z,k}\rangle^{1/2}\nonumber \\
 & =ia_{H,k}^{\dagger}e^{-i\theta}-ie^{i\theta}a_{H,k}
\end{align}
which defines $X_{k}$ and $P_{k}$ as quadrature phase amplitudes.
Hence, the fluctuation in $X_{k}$ ($P_{k}$) is due to $a_{H,k}$
only. Also, $[J_{x},J_{y}]=iJ_{z}$ ($\hbar=1$), and hence 
\begin{equation}
\langle[X_{k},P_{k}]\rangle{}_{I}=2\langle[J_{x,k},J_{y,k}]\rangle_{I}/\langle J_{z,k}\rangle=2i\langle J_{z,k}\rangle_{I}/\langle J_{z,k}\rangle\rightarrow2i
\end{equation}
and similarly for bipartitions of type $II$ and $III.$ This then
can be used to give the proof of Criterion 1. 
\begin{widetext}
\begin{eqnarray}
\frac{1}{\langle J_{z}\rangle}(\Delta^{2}u+\Delta^{2}v) & \geq & \frac{1}{2}\{P_{1}\{\left|g_{1}h_{1}\langle[X_{1},P_{1}]\rangle{}_{I}\right|+\left|g_{2}h_{2}\langle[X_{2},P_{2}]\rangle_{I}+g_{3}h_{3}\langle[X_{3},P_{3}]\rangle_{I}\right|\}\nonumber \\
 &  & +P_{2}\{\left|g_{2}h_{2}\langle[X_{2},P_{2}]\rangle_{II}\right|+\left|g_{1}h_{1}\langle[X_{1},P_{1}]\rangle_{II}+g_{3}h_{3}\langle[X_{3},P_{3}]\rangle_{II}\right|\}\nonumber \\
 &  & +P_{3}\{\left|g_{3}h_{3}\langle[X_{3},P_{3}]\rangle_{III}\right|+\left|g_{2}h_{2}\langle[X_{2},P_{2}]\rangle_{III}+g_{1}h_{1}\langle[X_{1},P_{1}]\rangle_{III}\right|\}\}\nonumber \\
 & {\normalcolor {\color{red}{\normalcolor =}}} & {\color{red}{\normalcolor P_{1}\{\left|g_{1}h_{1}\right|+\left|g_{2}h_{2}+g_{3}h_{3}\right|}{\normalcolor \}}}+P_{2}\{\left|g_{2}h_{2}\right|+\left|g_{1}h_{1}+g_{3}h_{3}\right|\}+P_{3}\{\left|g_{3}h_{3}\right|+\left|g_{2}h_{2}+g_{1}h_{1}\right|\}\nonumber \\
 & \geq & {\color{red}{\normalcolor min\{\left|g_{1}h_{1}\right|+\left|g_{2}h_{2}+g_{3}h_{3}\right|,}{\normalcolor \left|g_{2}h_{2}\right|+\left|g_{1}h_{1}+g_{3}h_{3}\right|,\left|g_{3}h_{3}\right|+\left|g_{2}h_{2}+g_{1}h_{1}\right|\}}}
\end{eqnarray}
\end{widetext}

The Criteria 1 and 3 are also applicable in Section V.B, where the
ensemble of atoms is considered to be large, and a large spin $J_{x}$
with minimal relative fluctuation ($\Delta^{2}J_{x}/\left|\langle J_{x}\rangle\right|\le1$)
is assumed. In Section V.C, the bipartite analysis of Eq. (\ref{eq:fadel_ent})
is unchanged. The use of Criterion 1 or 3 as in Eq. (\ref{eq:criterion3_BECsplitting})
is justified if $\Delta^{2}S_{x,Ak}/\left|\langle S_{x,Ak}\rangle\right|\le1$,
$\Delta^{2}S_{x,B}/\left|\langle S_{x,B}\rangle\right|\le1$, and
$\left|\langle S_{x,Ak}\rangle\right|$ and $\left|\langle S_{x,B}\rangle\right|$
are large.

\section{Conclusion}

In summary, we have derived several different criteria for certifying
full $N$-partite inseparability and genuine $N$-partite entanglement
using spin measurements. The criteria are inequalities expressed in
terms of variances of spin observables measured at each of the $N$
sites.

In Sections II and IV, we derive criteria based on the standard spin
uncertainty relation, involving $|\langle J_{z}\rangle|$. These criteria
are valid for any systems, provided at each site the outcomes are
reported faithfully, as results of accurately calibrated quantum measurements
\cite{Bancal_application_PRL2011,Opanchuk_PRA2014} . We present in
Section V three examples of application of these criteria. In these
examples, entanglement is created that can be detected using Stokes
or Schwinger operators defined at each site. These observables arise
naturally in atomic ensembles, where the creation and detection of
multi-partite entanglement is important for testing the quantum mechanics
of massive systems. The criteria we develop may be useful for this
purpose. In particular we specifically propose how to extend the experiments
of Julsgaard et al. \cite{Julsgaard:2001aa} and Fadel et al. \cite{Fadel409},
to generate three or more genuinely-entangled spatially-separated
ensembles of atoms. The experiment of Kunkel et al. \cite{Kunkel413}
succeeded in generating genuine $5$-partite entanglemen\textcolor{black}{t.}

Where Stokes operators are defined for atomic systems, it is possible
to introduce a normalization with respect to total atom number. This
concept was introduced by He et al. \textcolor{black}{\cite{He_EPR_strategies_PRL2011,He_planar_squeeze_NJP2012}}
and {\.{Z}}\textcolor{black}{ukowski et al. \cite{zukowski_PRA2016,zukowski_physicascripta2016,Zukowski_PRA2017,ryu_zukowski_arxiv2019}.
These authors show how the detection of entanglement and nonlocality
can be enhanced using such a normalization. It is likely that the
criteria derived in Sections II and IV may also be further improved
using t}his technique.

In Section III, we have outlined criteria derived from the planar
spin uncertainty relation $\Delta^{2}J_{x}+\Delta^{2}J_{y}\geq C_{J}$
valid for a system of fixed spin $J$. This is useful for states where
$\langle J_{z}\rangle=0$, such as the GHZ states. Such criteria were
developed previously for genuine tripartite steering. Although genuine
tripartite steering implies genuine tripartite entanglement, we have
extended the results of the earlier work by giving details of the
application of these criteria to certify the genuine tripartite entanglement
and the full tripartite inseparability of the GHZ and W states respectively.
While other methods exist to detect the genuine tripartite entanglement
of these states (for example \textit{\textcolor{black}{\emph{\cite{Collins_PRL2002,Toth_PRA2005,Korbicz_PRA2006}}}}),
the criteria we present in Section III are readily extended to higher
spin $J$. 
\begin{acknowledgments}
\textcolor{black}{This research has been supported by the Australian
Research Council Discovery Project Grants schemes under Grant DP180102470.
This work was performed in part at Aspen Center for Physics, which
is supported by National Science Foundation grant PHY-1607611.} We
thank Manuel Gessner and Matteo Fadel for invaluable discussions and
for pointing out the problems with Eq. (\ref{eq:sum_arbitrary_bipartition})
in the initial version of our manuscript. 
\end{acknowledgments}

\section*{Appendix}

\subsection*{1. Lower bound of the sum inequality for an arbitrary bipartition}

We derive the inequality in Eq. (\ref{eq:varu_plus_varv_arbitrary})
for an arbitrary pure state bipartition $\rho_{k}^{\zeta}\rho_{lm}^{\zeta}$.
\begin{widetext}
\begin{align}
\Delta^{2}u_{\zeta}+\Delta^{2}v_{\zeta} & =\Delta^{2}\left(h_{k}J_{x,k}\right)_{\zeta}+\Delta^{2}\left(h_{l}J_{x,l}+h_{m}J_{x,m}\right)_{\zeta}+\Delta^{2}\left(g_{k}J_{y,k}\right)_{\zeta}+\Delta^{2}\left(g_{l}J_{y,l}+g_{m}J_{y,m}\right)_{\zeta}\nonumber \\
 & \geq\left|g_{k}h_{k}\langle\left[J_{x,k},J_{y,k}\right]\rangle_{\zeta}\right|+\left|g_{l}h_{l}\langle\left[J_{x,l},J_{y,l}\right]\rangle_{\zeta}+g_{m}h_{m}\langle\left[J_{x,m},J_{y,m}\right]\rangle_{\zeta}\right|\nonumber \\
 & =\left|g_{k}h_{k}\langle J_{z,k}\rangle_{\zeta}\right|+\left|g_{l}h_{l}\langle J_{z,l}\rangle_{\zeta}+g_{m}h_{m}\langle J_{z,m}\rangle_{\zeta}\right|\,.\label{eq:sum_arbitrary_bipartition}
\end{align}
\end{widetext}

Here the subscript $\zeta$ denotes that the statistical moments are
calculated with respect to the particular state with bipartition $\rho_{k}^{\zeta}\rho_{lm}^{\zeta}$.
The uncertainty relation $\Delta^{2}\left(hJ_{x}\right)_{\zeta}+\Delta^{2}\left(gJ_{y}\right)_{\zeta}\geq|\langle gh\left[J_{x},J_{y}\right]\rangle|_{\zeta}$
is used to obtain the first inequality in Eq. (\ref{eq:sum_arbitrary_bipartition}).
The spin commutation relation $\left[J_{x},J_{y}\right]=iJ_{z}$ is
used in the last line.

\subsection*{2. Lower bound of the product inequality for an arbitrary bipartition
\label{subsec:1}}

We derive the inequality in Eq. (\ref{eq:arbitrary_bipartition-1})
for an arbitrary bipartition $\rho_{k}^{\zeta}\rho_{lm}^{\zeta}$.
\begin{widetext}
\begin{align}
\Delta^{2}u_{\zeta}\Delta^{2}v_{\zeta} & =\left[\Delta^{2}\left(h_{k}J_{x,k}\right)_{\zeta}+\Delta^{2}\left(h_{l}J_{x,l}+h_{m}J_{x,m}\right)_{\zeta}\right]\left[\Delta^{2}\left(g_{k}J_{y,k}\right)_{\zeta}+\Delta^{2}\left(g_{l}J_{y,l}+g_{m}J_{y,m}\right)_{\zeta}\right]\nonumber \\
 & =\Delta^{2}\left(h_{k}J_{x,k}\right)_{\zeta}\Delta^{2}\left(g_{k}J_{y,k}\right)_{\zeta}+\Delta^{2}\left(h_{l}J_{x,l}+h_{m}J_{x,m}\right)_{\zeta}\Delta^{2}\left(g_{l}J_{y,l}+g_{m}J_{y,m}\right)_{\zeta}\nonumber \\
 & +\Delta^{2}\left(h_{l}J_{x,l}+h_{m}J_{x,m}\right)_{\zeta}\Delta^{2}\left(g_{k}J_{y,k}\right)_{\zeta}+\Delta^{2}\left(h_{k}J_{x,k}\right)_{\zeta}\Delta^{2}\left(g_{l}J_{y,l}+g_{m}J_{y,m}\right)_{\zeta}\nonumber \\
 & \geq\Delta^{2}\left(h_{k}J_{x,k}\right)_{\zeta}\Delta^{2}\left(g_{k}J_{y,k}\right)_{\zeta}+\Delta^{2}\left(h_{l}J_{x,l}+h_{m}J_{x,m}\right)_{\zeta}\Delta^{2}\left(g_{l}J_{y,l}+g_{m}J_{y,m}\right)_{\zeta}\nonumber \\
 & +2\Delta\left(h_{l}J_{x,l}+h_{m}J_{x,m}\right)_{\zeta}\Delta\left(g_{k}J_{y,k}\right)_{\zeta}\Delta\left(h_{k}J_{x,k}\right)_{\zeta}\Delta\left(g_{l}J_{y,l}+g_{m}J_{y,m}\right)_{\zeta}\nonumber \\
 & =\left[\Delta\left(h_{k}J_{x,k}\right)_{\zeta}\Delta\left(g_{k}J_{y,k}\right)_{\zeta}+\Delta\left(h_{l}J_{x,l}+h_{m}J_{x,m}\right)_{\zeta}\Delta\left(g_{l}J_{y,l}+g_{m}J_{y,m}\right)_{\zeta}\right]^{2}\nonumber \\
 & \geq\left[\frac{\left|g_{k}h_{k}\langle J_{z,k}\rangle_{\zeta}\right|}{2}+\frac{\left|g_{l}h_{l}\langle J_{z,l}\rangle_{\zeta}+g_{m}h_{m}\langle J_{z,m}\rangle_{\zeta}\right|}{2}\right]^{2}\,.\label{eq:varuvarv_inequality}
\end{align}
\end{widetext}

In going from the second equality to the first inequality, the inequality
for two real numbers $x$ and $y$, $x^{2}+y^{2}\geq2xy$, is used.
The uncertainty relation in the final line is $\Delta\left(hJ_{x}\right)_{\zeta}\Delta\left(gJ_{y}\right)_{\zeta}\geq\langle\left|gh\left[J_{x},J_{y}\right]\right|\rangle_{\zeta}/2$.

\subsection*{}

\vskip -0.4in 

\subsection*{3. Generating genuine tripartite entangled states using 3 beam splitters
and one squeezed input}

Here we consider the configuration of Fig. \ref{fig:tripartite_ent_BS}.
The output mode operators $a_{\text{out}}$, $b_{\text{out}}$ and
$c_{\text{out}}$ are 
\begin{align}
a_{\text{out}} & =\frac{1}{\sqrt{3}}a_{\text{in}}+\sqrt{\frac{2}{3}}b_{\text{in}}\nonumber \\
b_{\text{out}} & =\frac{1}{\sqrt{2}}\left(\sqrt{\frac{2}{3}}a_{\text{in}}-\frac{1}{\sqrt{3}}b_{\text{in}}\right)+\frac{1}{\sqrt{2}}c_{\text{in}}\nonumber \\
c_{\text{out}} & =\frac{1}{\sqrt{2}}\left(\sqrt{\frac{2}{3}}a_{\text{in}}-\frac{1}{\sqrt{3}}b_{\text{in}}\right)-\frac{1}{\sqrt{2}}c_{\text{in}}\,.\label{eq:mode_outputs}
\end{align}
Now, we consider $X_{a,\text{out}}-1/2\left(X_{b,\text{out}}+X_{c,\text{out}}\right)$
and $P_{a,\text{out}}+1/2\left(P_{b,\text{out}}+P_{c,\text{out}}\right)$.
Their variances are then 
\begin{align}
\Delta^{2}\!\left[\!X_{a,\text{out}}\!-\!\frac{\left(X_{b,\text{out}}+X_{c,\text{out}}\right)}{2}\right]\! & =\frac{3}{2}\Delta^{2}X_{b,\text{in}}=\frac{3}{2}\nonumber \\
\Delta^{2}\!\left[P_{a,\text{out}}+\frac{\left(P_{b,\text{out}}+P_{c,\text{out}}\right)}{2}\right]\! & =\frac{2}{3}\Delta^{2}P_{a,\text{in}}+\frac{1}{6}\Delta^{2}P_{b,\text{in}}\nonumber \\
 & =\frac{2}{3}e^{-2r}+\frac{1}{6}\label{eq:output_variances}
\end{align}
and their sum is \textcolor{black}{\small{}{} 
\begin{gather}
\Delta^{2}\!\left[\!X_{a,\text{out}}\!-\!\frac{\left(X_{b,\text{out}}+X_{c,\text{out}}\right)}{2}\!\right]\!+\!\Delta^{2}\!\left[\!P_{a,\text{out}}\!+\!\frac{\left(P_{b,\text{out}}+P_{c,\text{out}}\right)}{2}\!\right]\nonumber \\
=\frac{10}{6}+\frac{2}{3}e^{-2r}\,,\label{eq:output_sum}
\end{gather}
}{\small{}giving a minimum of $10/6=1.6667$ for large squeezing parameter
$r$. The sum inequality for those variances is }\textcolor{black}{\small{}$\Delta^{2}\left[X_{a,\text{out}}-1/2\left(X_{b,\text{out}}+X_{c,\text{out}}\right)\right]+\Delta^{2}\left[P_{a,\text{out}}+1/2\left(P_{b,\text{out}}+P_{c,\text{out}}\right)\right]\geq2$
, as shown in Criterion 5 of Ref. \cite{PhysRevA.90.062337} with
the gains $h=-1/2$ and $g=1/2$. This inequality is violated }{\small{}and
hence the final output state in Fig. \ref{fig:tripartite_ent_BS}
is genuinely tripartite entangled. }\textcolor{black}{\small{}We can
also consider the input to be squeezed along $X$, in which case the
gains $g$ and $h$ will have opposite sign.}{\small\par}

\subsection*{4. Generating genuine four-partite entangled states using $4$ beam
splitters and one squeezed input}

\textcolor{black}{Here we consider the configuration of Fig. \ref{fig:fourpartite_ent_BS}.
The output mode operators $a_{\text{out}}$, $b_{\text{out}}$ , $c_{\text{out}}$
and $d_{\text{out}}$ are 
\begin{align}
a_{\text{out}} & =\frac{1}{\sqrt{4}}a_{\text{in}}+\sqrt{\frac{3}{4}}b_{\text{in}}\nonumber \\
b_{\text{out}} & =\frac{1}{\sqrt{3}}\left(\sqrt{\frac{3}{4}}a_{\text{in}}-\frac{1}{\sqrt{4}}b_{\text{in}}\right)+\sqrt{\frac{2}{3}}c_{\text{in}}\nonumber \\
c_{\text{out}} & =\frac{1}{\sqrt{3}}\left(\sqrt{\frac{3}{4}}a_{\text{in}}-\frac{1}{\sqrt{4}}b_{\text{in}}\right)-\frac{1}{\sqrt{6}}c_{\text{in}}+\frac{1}{\sqrt{2}}d_{\text{in}}\nonumber \\
d_{\text{out}} & =\frac{1}{\sqrt{3}}\left(\sqrt{\frac{3}{4}}a_{\text{in}}-\frac{1}{\sqrt{4}}b_{\text{in}}\right)-\frac{1}{\sqrt{6}}c_{\text{in}}-\frac{1}{\sqrt{2}}d_{\text{in}}\,.\label{eq:4_partite_mode_outputs}
\end{align}
Now, we consider $X_{a,\text{out}}-1/3\left(X_{b,\text{out}}+X_{c,\text{out}}+X_{d,\text{out}}\right)$
and $P_{a,\text{out}}+1/3\left(P_{b,\text{out}}+P_{c,\text{out}}+P_{d,\text{out}}\right)$.
Their variances are then 
\begin{align}
\Delta^{2}\!\left[\!X_{a,\text{out}}\!-\!\frac{\left(X_{b,\text{out}}+X_{c,\text{out}}+X_{d,\text{out}}\right)}{3}\!\right]\!\! & =\frac{4}{3}\Delta^{2}X_{b,\text{in}}=\frac{4}{3}\nonumber \\
\!\Delta^{2}\left[P_{a,\text{out}}+\frac{\left(P_{b,\text{out}}+P_{c,\text{out}}+P_{d,\text{out}}\right)}{3}\right]\!\! & =\!\!\Delta^{2}\!P_{a,\text{in}}\!+\!\frac{1}{3}\Delta^{2}\!P_{b,\text{in}}\nonumber \\
 & =e^{-2r}+\frac{1}{3}\label{eq:output_variances-1}
\end{align}
and their sum is $5/3+e^{-2r}$, giving a minimum of $5/3=1.6667$
for large squeezing parameter $r$. The sum inequality for those variances
is $\Delta^{2}\left[X_{a,\text{out}}-1/3\left(X_{b,\text{out}}+X_{c,\text{out}}+X_{d,\text{out}}\right)\right]+\Delta^{2}\left[P_{a,\text{out}}+1/3\left(P_{b,\text{out}}+P_{c,\text{out}}+P_{d,\text{out}}\right)\right]\geq16/9$,
as shown in Criterion 8 of Ref. \cite{PhysRevA.90.062337} for $N=4$
and with the gains $h=-1/3$ and $g=1/3$. This inequality is violated
and hence the final output state in Fig. \ref{fig:fourpartite_ent_BS}
is genuinely four-partite entangled. We note we can also consider
the input to be squeezed along $X$, in which case the gains $g$
and $h$ will have opposite sign.}

\subsection*{5. Beam splitter operation as a model for splitting BEC clouds}

\textcolor{black}{We define the mode operators $a_{+}=\left(a_{+1}-ia_{+2}\right)/\sqrt{2}$
and $a_{-}=\left(a_{-1}-ia_{-2}\right)/\sqrt{2}$, and their corresponding
auxiliary mode operators $a_{\text{vac}+}=\left(a_{+1}-ia_{+2}\right)/\sqrt{2}$
and $a_{\text{vac}-}=\left(a_{+1}-ia_{+2}\right)/\sqrt{2}$. This
allows us to model the splitting of a BEC cloud with the beam splitter
operations where the mode operators $a_{+},a_{\text{vac}+}$ are the
inputs of a beam splitter. Since the different spin species does not
interact, the mode operators $a_{-},a_{\text{vac}-}$ are also the
inputs of a beam splitter and are split independently of the other
spin species. With these mode operators, the Schwinger spin operators
after splitting are \vfill{}
\begin{align}
S_{z,A1} & =\frac{1}{2}\left(a_{+1}^{\dagger}a_{+1}-a_{-1}^{\dagger}a_{-1}\right)\nonumber \\
 & =\frac{1}{4}\left(a_{+}^{\dagger}a_{+}-a_{-}^{\dagger}a_{-}\right)+F\left(a_{\text{vac}+},a_{\text{vac}-}\right)\label{eq:S_zA1}\\
S_{z,A2} & =\frac{1}{2}\left(a_{+2}^{\dagger}a_{+2}-a_{-2}^{\dagger}a_{-2}\right)\nonumber \\
 & =\frac{1}{4}\left(a_{+}^{\dagger}a_{+}-a_{-}^{\dagger}a_{-}\right)+G\left(a_{\text{vac}+},a_{\text{vac}-}\right)\,.\label{eq:S_zA2}
\end{align}
Here we take the orientation of $x,y,z$ so that $S_{z}$ corresponds
to the number difference. $S_{z,A1}$ and $S_{z,A2}$ are the Schwinger
spin operators along the $z$-axis for clouds $A1$ and $A2$ respectively,
$F$ and $G$ are terms containing $a_{\text{vac}+},a_{\text{vac}+}^{\dagger},a_{\text{vac}-},a_{\text{vac}-}^{\dagger}$.
Similar Schwinger spin operators along the $x$ and $y$-axes have
the same expressions as Eqs. (\ref{eq:S_zA1}) and (\ref{eq:S_zA2})
but the spin up and down are relative to their respective axis. From
Eqs. (\ref{eq:S_zA1}) and (\ref{eq:S_zA2}), we see that $S_{z,A1}+S_{z,A2}=S_{z,A}+F+G\approx S_{z,A}$.
Here we assume the terms $F$ and $G$ involving the incoming unoccupied
modes can be neglected in the calculation of the variances, relative
to the leading terms which come from the incoming modes with a high
occupation (the number of atoms being assumed large). Using a similar
argument, we consider $S_{\theta}=\frac{1}{2}\left(a_{+}^{\dagger}a_{-}e^{i\theta}+a_{-}^{\dagger}a_{+}e^{-i\theta}\right)\,.$
\begin{align*}
S_{\theta,A1} & =\frac{1}{2}\left(a_{+1}^{\dagger}a_{-1}e^{i\theta}+a_{-1}^{\dagger}a_{+1}e^{-i\theta}\right)\\
 & =\frac{1}{4}\left(a_{+}^{\dagger}a_{-}e^{i\theta}+a_{-}^{\dagger}a_{+}e^{-i\theta}\right)+F(a_{\text{vac}+},a_{\text{vac}-})\\
\\
S_{\theta,A2} & =\frac{1}{2}\left(a_{+2}^{\dagger}a_{-2}e^{i\theta}+a_{-2}^{\dagger}a_{+2}e^{-i\theta}\right)\\
 & =\frac{1}{4}\left(a_{+}^{\dagger}a_{-}e^{i\theta}+a_{-}^{\dagger}a_{+}e^{-i\theta}\right)+G(a_{\text{vac}+},a_{\text{vac}-})
\end{align*}
So, for large numbers of atoms, $S_{y,A1}+S_{y,A2}=S_{y,A}+F+G\approx S_{y,A}$
and similarly, $S_{x,A1}+S_{x,A2}\approx S_{x,A}$.}

\bibliographystyle{apsrev4-1}
\bibliography{Genuine_multipartite_entanglement_spin_submission-arxiv-Sept2020}

\end{document}